\documentclass[12pt]{article}
\usepackage{amsmath}

\usepackage[english]{babel}
\usepackage[ansinew]{inputenc}
\usepackage{epsfig}
\usepackage{color,graphicx}
\usepackage{wrapfig}
\usepackage{amssymb,amsthm,amsmath}
\usepackage{dsfont}
\usepackage{natbib}
\usepackage{abstract}
\usepackage{enumerate}
\usepackage{enumitem}

\renewcommand{\baselinestretch}{1}

\newtheorem{lemma}{Lemma}[]

\newtheorem{proposition}{Proposition}[]

\newtheorem*{thma}{Proof}

\def\ds{\displaystyle}

\begin{document}

\title{\bf Exact Bayesian inference in spatiotemporal Cox processes driven by multivariate Gaussian processes}
\author{\bf{Flávio B. Gonçalves$^{a}$, Dani Gamerman$^{b}$}}
\date{}

\maketitle

\begin{center}
{\footnotesize $^a$ Departamento de Estatística, Universidade Federal de Minas Gerais, Brazil\\
$^b$ Departamento de Métodos Estatísticos, Universidade Federal do Rio de Janeiro, Brazil}
\end{center}

\footnotetext[1]{Published in the Journal of the Royal Statistical Society - Series B.}

\footnotetext[2]{Address: Av. Antônio Carlos, 6627 - DEST/ICEx/UFMG -
Belo Horizonte, Minas Gerais, 31270-901, Brazil. E-mail: fbgoncalves@est.ufmg.br}

\renewcommand{\abstractname}{Abstract}
\begin{abstract}

In this paper we present a novel inference methodology to perform Bayesian inference
for spatiotemporal Cox processes where the intensity function depends on a multivariate Gaussian process.
Dynamic Gaussian processes are introduced to allow for evolution of the intensity function over discrete time.
The novelty of the method lies on the fact that no discretisation error is involved despite the non-tractability
of the likelihood function and infinite dimensionality of the problem.
The method is based on a Markov chain Monte Carlo algorithm that samples from the joint posterior distribution of
the parameters and latent variables of the model. The models are defined in a general and flexible way but they
are amenable to direct sampling from the relevant distributions,
due to careful characterisation of its components.
The models also allow for the inclusion of regression covariates and/or temporal components to explain the variability of the intensity function.
These components may be subject to relevant interaction with space and/or time.
Real and simulated examples illustrate the methodology, followed by concluding remarks.


\end{abstract}


\section{Introduction}

A Cox process is an inhomogeneous Poisson process where the intensity function evolves stochastically. It is also referred to as doubly stochastic process. Cox processes \citep{Cox2} have been extensively used in a variety of areas to model point process phenomena.
Effects in Cox processes may present spatio-temporal variation to reflect the possibility of interaction
between space-time and other model components. They can be traced back to log-Gaussian Cox processes
\citep{moller}, where a Gaussian Process (GP) representation is used for the log-intensity \citep[see also][and references within]{diggle2}.

The application of (GP driven) Cox processes is closely related to two main problems: simulation and inference. These are hard problems due to the infinite dimensionality of the process and the intractability of the likelihood
function. Simulation is one of the main tools to tackle the inference problem which primarily consists of estimating the unknown intensity function (IF) and potential unknown parameters. However, prediction is often a concern, i.e., what should one expect in a future realisation of the same phenomenon.

Solutions for the inference problem have required, until recently, the use of discrete approximations \citep[see, for example,][]{moller,diggle,dani2}. These represent a considerable source of error and, therefore, ought to be used with care \citep[see][]{simpson}. Approximated (discretisation-based) methods have as its main disadvantages: $i)$ producing biased results; $ii)$ it is hard (sometimes impossible) to quantify the bias (error) involved; $iii)$ the level of discretisation required to obtain good (close enough to the exact) results is unknown and case-specific;  $iv)$ it is not always clear to which limit these approaches converge to.
This motivates the development of exact methodologies, i.e. free from discretisation errors and helps to understand its advantages.

The importance of a given exact approach basically relies on its applicability in terms of modelling flexibility and computational cost. Exact solutions for inference on infinite-dimensional processes with intractable likelihood can be found, for example, in \citet{bpr06a}, \citet{sermai} and \citet{flavio3}. In this paper, the term exact refers to the fact that no discretisation-based approximation is used. In particular, the methodology proposed here has MCMC error (Markov chain convergence + Monte Carlo) as the only source of inaccuracy, which is generally well understood and controlled.

One non-parametric exact approach to the analysis of spatial point patterns was proposed in \citet{adams}. They consider a univariate Gaussian process to describe the IF dynamics and an augmented model for the data and latent variables that simplifies the likelihood function. Theoretical concern about \citet{adams}'s algorithm and empirical evidence to support it are provided in Section 5 of the Appendix. Another non-parametric exact approach was adopted in \citet{sanso2}, where a particular factorisation of the IF was proposed and Dirichlet processes priors were used. Their work was extended to the spatio-temporal context by \citet{sanso1}.

The aim of this work is to propose an exact inference methodology for spatio-temporal Cox processes in which the intensity function dynamics is driven by a Gaussian process. The exactness feature stems from an augmented model approach as in \citet{adams}. However, we generalise their point pattern models by firstly considering spatio-temporal models and, secondly, by using multivariate (possibly dynamic) Gaussian processes to allow the inclusion of different model components (regression and temporal effects) in a flexible manner.
Space and time may be considered continuous or discrete. In this paper we ultimately consider the general formulation of continuous space and discrete time. This is actually the most general formulation, since its continuous time version can be seen as a continuous space process where time is one of the dimensions.

Our methodology also introduces a particularly suited MCMC algorithm that enables direct simulation from the full conditional distributions of the Gaussian process and of other relevant latent variables. Moreover, estimation of (possibly intractable) functionals of the IF and prediction based on the output of the MCMC is straightforward. We also provide formal proofs of the validity of the MCMC algorithm.

We believe our methodology is the first valid exact alternative for inference in GP-driven Cox processes. It is robust under model complexity/specification, i.e. it is valid and has the same general formulation for any model where the prior measure on the main component of the IF  is a Gaussian process - uni or multivariate, space or space-time, with or without covariates, etc. Furthermore, its complexity is mitigated by the good convergence properties of the proposed MCMC algorithm and computational time optimisation strategies based on matrix algebra results. Finally, we present two real examples where we compare our methodology to a standard approximated one and demonstrates its applicability to reasonably sized problems.

This paper is organised as follows. Section \ref{secmod} presents the class of models to be considered and the augmented model to derive the MCMC.
Section \ref{secinf} presents the general Bayesian approach and addresses some identifiability and implementation issues. Section \ref{seccomp} describes the MCMC algorithm for the spatial model and Section \ref{stmsec} presents its extension to the spatio-temporal case. Section \ref{secsim} presents two real data examples to illustrate the proposed methodology. Final remarks and possible directions for future work are presented in Section \ref{secconc}. The Appendix presents the aforementioned analysis of \citet{adams}'s algorithm as well as a collection of simulated data analysis, proofs for the main results and other practical aspects of our methodology.


\section{Model specification}\label{secmod}

In this section we present the complete probabilistic model for spatiotemporal point processes with Gaussian process driven intensities.
We break the presentation in parts considering the different levels and generalisations of the model.

\subsection{The general Cox process model}\label{secmod1}

We consider a Poisson process (PP) $Y=\{Y_t;\;t\in\mathcal{T}\}$ in $S\times \mathcal{T}$, where $S$ is some compact region in $\mathds{R}^d$ and $\mathcal{T}$ is a finite set of $\mathds{N}$. It can be seen as a Poisson process in a region $S$ that evolves in time. We assume the Poisson process has an intensity function $\lambda_t(s):S\times \mathcal{T}\rightarrow\mathds{R}^+$.

We assume that the IF is a function of a (multivariate spatiotemporal) Gaussian process and covariates.
A Gaussian process $\beta$ is a stochastic process in some space such that the joint distribution of any finite collection of points in this space is Gaussian. This space may be defined so that we have spatial or spatiotemporal processes.

A detailed presentation of (dynamic) Gaussian processes is given in Section \ref{ssgp}. For now, let $\beta:=\{\beta_0,\beta_1,\ldots,\beta_q\}$ be a collection of $q+1$ independent GP's (a multivariate GP) in $S\times \mathcal{T}$, for $\mathcal{T}=\{0,\ldots,T\}$. We assume the following model for $Y$:

\begin{eqnarray}
  (Y_t|\lambda_{S,t}) &\sim& PP(\lambda_{S,t}),\;\;\forall\;t\in\mathcal{T}, \label{gpif1}\\
  \lambda_t(s)&=&\lambda_{t}^*\Phi(f(\beta_t(s),W_t(s))),\;\;\forall\;t\in\mathcal{T}, \label{gpif2}\\
  (\beta|\theta) &\sim& GP_{\theta}, \label{gpif3}\\
  (\lambda_{\mathcal{T}}^*,\theta)&\sim& prior, \label{gpif4}
\end{eqnarray}
where $\lambda_{\mathcal{T}}^*=(\lambda_{0}^*,\ldots,\lambda_{T}^*)$, $\Phi$ is the distribution function of the standard Gaussian distribution, $GP_{\theta}$ is a Gaussian process indexed by (unknown) parameters $\theta$ and $W=\{W_t(s)\}$ is a set of covariates. We assume $f$ to be linear in the coordinates of $\beta$ and write $W$ in a way such that $f(\beta_t(s),W_t(s))=W_t(s)\beta_t(s)$. Any distribution function of a continuous r.v. or any general bounded function may be used instead of $\Phi$. For example, a common choice is the logistic function, which is used by \citet{adams} and is very similar to $\Phi$ (the largest difference is 0.0095). The particular choice in (\ref{gpif2}) contributes to the construction of the MCMC algorithm as discussed in Section \ref{seccomp}.
These links from the GP to the IF are bounded to $[0,1]$ in contrast with the unbounded exponential link of the log-Gaussian Cox processes (LGCP) proposed by
\citet{moller}.
We are interested in estimating not only the overall rate $\lambda_{S,\mathcal{T}}$ but also the univariate GP's, separately, given their possibly meaningful interpretation.

Parameter $\lambda_{t}^*$ ought to represent the supremum of the intensity function
at time $t$. One extreme possibility is to assume $\lambda_{t}^* = \lambda^*$,
which is a reasonable assumption in the case processes whose maximum intensity is time invariant. In this case, a common choice for
the prior distribution of $\lambda^*$ is $\mathcal{G}(\alpha_{\lambda},\beta_{\lambda})$ - a Gamma distribution. At the other extreme,
unrelated parameters vary independently over time according to
independent $\mathcal{G}(\alpha_{\lambda_t},\beta_{\lambda_t})$ prior distributions.
In between them, models allow for temporal dependence between
(successive) $\lambda_{t}^*$'s. One such formulation with attractive features is briefly
described in Section \ref{stmsec}.

The Gaussian processes may represent a number of relevant model features such as the effect of covariates
and/or spatiotemporal components such as the spatially-varying trend components or seasonality of the baseline intensity. They may also be space and/or time invariant. One common example is the model with $q$ covariates:
\begin{equation}\label{gpif3}
\ds f(\beta_t(s),W_t(s))=\beta_{0,t}(s)+\beta_{1,t}(s)W_{1,t}(s)+\ldots+\beta_{q,t}(s)W_{q,t}(s).
\end{equation}
This approach allows the use of extra prior information through covariates. The spatiotemporal variation of the effects is particularly relevant in applications where covariates present significant
interaction with space and/or time. Examples are provided in \citet{pinto}.

The first advantage of the formulation in (\ref{gpif1})-(\ref{gpif4}) is that it allows exact simulation of data from the model which is the key to develop exact inference methods. Exact simulation of the model is based on a key result from Poisson processes called Poisson thinning. This is a variant of rejection sampling for point processes proposed by \citet{lewis} and is given in Algorithm 1 below:\\
\\{\renewcommand{\baselinestretch}{0.5}\scriptsize
\begin{tabular}[!]{|l|}
\hline\\
\parbox[!]{14cm}{
\texttt{
{\bf Algorithm 1}
\begin{enumerate}
\setlength\itemsep{-0.5em}
\item simulate a Poisson process $\ds(s_1,\ldots,s_K)$ with constant intensity function $\ds \lambda_{t}^*$ on $S$:
\begin{enumerate}
\setlength\itemsep{-0.5em}
  \item simulate $K_t\sim Poisson(\lambda_{t}^*\mu(S))$, where $\mu(S)$ is the volume of $S$;
  \item distribute the $K_t$ points uniformly on $S$.
\end{enumerate}
\item simulate $\beta_t$ and observe $W_t$ at points $\{s_1,\ldots,s_K\}$;
\item keep each of the $K_t$ points with probability $\ds \lambda_t(s_k)/\lambda_{t}^*$;
\item output the points kept on the previous step.
\end{enumerate}
}}\\  \hline
\end{tabular}}\\
\\
To simulate the process at additional times $t\in \mathcal{T}$, it is enough to perform the algorithm above for each $t$ and simulate the Gaussian process conditional on the points previously simulated. The idea of Poisson thinning is applied in related contexts by \citet{flavio2}.

\subsection{The augmented model}

Performing exact inference for Cox processes is a challenging problem mainly because of the intractability of the likelihood function, which is given by
\begin{equation}\label{PPL}
L(\lambda_{S,\mathcal{T}},y)=\exp\left\{-\sum_{t=0}^T\int_{S}\lambda_t(s)ds\right\}\prod_{t=1}^T\prod_{n=1}^{N_t}\lambda_t(s_{t,n}),
\end{equation}
where $s_{n,t}$ is the location of the $n$-th event of $Y_t$.

The crucial step to develop exact methods is to avoid dealing with the likelihood above. One possible solution is to define an augmented model for $Y$ and some additional variable $X$, such that the joint (pseudo-)likelihood based on $(X,Y)$ is tractable. This poses the problem as a missing data problem (where inference is based on the joint (pseudo-)likelihood of the data and some latent variable) and allows us to use standard methods. The augmented model is constructed based on the Poisson thinning presented in Algorithm 1. The same strategy is employed in \citet{adams} to circumvent the non-tractability problem.

Firstly, define $X=\{X_t;\;t\in\mathcal{T}\}$ where each $X_t$ is a homogeneous PP with intensity $\lambda_{t}^*$ on $S$ and the $X_t$'s are mutually independent.
Now let $\{s_{t,k}\}_{k=1}^{K_t}$ be the locations of the $K_t$ events of $X_t$.
We also define $T$ vectors $Z_t$, $t=1,\ldots,T$, with each coordinate taking values in $\{0,1\}$ such that $(Z_t|X,\beta_{K_t},W_{K_t})$ is a random vector $(Z_{t,1},\ldots,Z_{t,K_t})$, where the $Z_{t,k}$'s are all independent with $Z_{t,k}\sim Ber(\Phi(W_t(s_{t,k})\beta_t(s_{t,k})))$ and $(\beta_{K_t},W_{K_t})$ is $(\beta,W)$ at the points from $X_t$. Finally, define $Y_t=h(Z_t,X_t)$ as the non-zero coordinates of the vector $(Z_{t,1}s_{t,1},\ldots,Z_{t,K_t}s_{t,K_t})$, which leads to the model (\ref{gpif1}).

Namely, the augmented model defines $Y$ as the events remaining from performing the Poisson thinning to a PP $X$. It is important to note, however, that only $Y$ is observed. We define $\{s_{t,n}\}_{n=1}^{N_t}$ as the $N_t$ events of $Y_t$ and $\{s_{t,m}\}_{m=1}^{K_t-N_t}$ as the $M_t:= K_t-N_t$ thinned events. Most importantly, this approach leads to a tractable likelihood when the joint distribution of $X$ and $Y$ is considered, as it is shown in Section \ref{secspace}.

The spatial model is a particular case where $T=1$ which implies that $X$ and $Y$ are Poisson processes on $S$ with intensity functions $\lambda^*$ and $\lambda(s)$, respectively. We observe $\{s_n\}_{n=1}^{N}$ from $Y$ and simplify the notation above accordingly. Note that the spatial model for unidimensional $S$ is generally seen as the commonly used Cox process in (continuous) time.

\subsection{Dynamic Gaussian processes}\label{ssgp}

Gaussian processes are a very flexible component to handle spatial variation,
specially when smooth processes are expected. We say that $\beta$ follows a stationary Gaussian
process in $S$ if $\beta ( s ) \sim N( \mu , \sigma^2 )$ and $Cov ( \beta ( s ) ,
\beta ( s' ) ) = h ( s , s' )$, for $s , s' \in S$, constants $\mu$ and $\sigma^2$
and a (almost everywhere) differentiable function $h$. Further simplification is obtained if isotropy
can be assumed, leading to $h ( s , s' ) = \rho ( | s - s' | )$. In this case,
the process is denoted by $\beta \sim GP ( \mu , \sigma^2 , \rho )$ and $\rho$ is
referred to as the correlation function.

Typical choices for $h$ belong to the $\gamma$-exponential family of covariance functions:
\begin{equation}\label{gecf}
h ( s , s' )=\sigma^2\exp\left\{-1/(2\tau^2)|s-s'|^{\gamma}\right\},\;\;\;0<\gamma\leq2.
\end{equation}
The case where $\gamma>1$ leads to almost-surely differentiable paths (surfaces).

GP's can be extended in many directions. The most important ones here are
extensions to handle multivariate GP's and extensions to cope with space and time.
There are a number of different ways to allow for multivariate responses. The main
ones are reviewed in \citet{dani5} and include independent GP's,
dependent processes with a common correlation function, or linear mixtures of
independent GP's, the coregionalisation models described by
\citet{Wack}. Similar ideas were also developed within the machine learning
literature as multiple output models \citep[see, for example,][]{BF04}. A review by
\citet{ARL} discusses the relation between coregionalisation and multiple output models.

Extensions to cope with space and discrete time were considered by \citet{dani6} \citep[see also][]{WC99}.
A process $\beta$ follows a dynamic
Gaussian process in discrete time if it can be described by a difference equation
\begin{equation}\label{DGP}
\beta_{t'}(\cdot) = G_{t',t}\beta_t(\cdot) + w_{t',t}(\cdot) \quad , \quad w_{t',t}\sim GP,
\end{equation}
where the multivariate Gaussian process disturbances $w_{t',t}(\cdot)$ are zero mean and time-independent;
they are also taken as identically distributed in the equidistant case $t'= t+1$.
The law of the process is completed with a Gaussian process specification for $\beta_0(\cdot)$.
Similar processes were proposed in continuous time by \citet{diggle}.

A number of options are available for the temporal transition matrix $G$, including the
identity matrix. If additionally the disturbance processes $w$ consist of independent GP's
then the resulting process consist of independent univariate dynamic GP's.
\citet{Gam10} presents some alternatives to model trend and seasonality of the IF.

It is also possible to consider the use of non-spatiotemporal covariates to explain the intensity function variation. This approach, however, requires adaptations to the original model and can be found in \citet{pinto}.


\section{Inference for the spatial model}\label{secinf}

We now focus on the inference problem of estimating the intensity function $\lambda_{S,\mathcal{T}}$, parameter $\lambda^*$ and potentially unknown parameters $\theta$ from the Gaussian process, based on observations from the Poisson process $Y$. We shall also discuss how to make prediction. In order to make the presentation of the methodology as clear as possible, we consider first the (purely) spatial process and then the generalisation for the spatiotemporal case.

\subsection{Posterior distribution}\label{secspace}

Let $\{s_k\}_{k=1}^{K}$ and $\{s_n\}_{n=1}^{N}$ be the events from $X$ and $Y$, respectively, and $\{s_m\}_{m=1}^{K-N}$ be the thinned events.
Furthermore, $\beta_K$, $\beta_N$, and $\beta_M$ are $\beta$ at $\{s_k\}_{k=1}^{K}$, $\{s_n\}_{n=1}^{N}$ and $\{s_m\}_{m=1}^{K-N}$, respectively, and $W_K$, $W_N$ and $W_M$ are the respective subsets of $W$. Note that $\ds \{s_k\}_{k=1}^{K}=\left(\{s_n\}_{n=1}^{N}\bigcup\{s_m\}_{m=1}^{K-N}\right)$ and $\beta_K=(\beta_N,\beta_M)$.
Defining $\beta_{-K}$ as the Gaussian process in $S\setminus\{s_k\}_{k=1}^{K}$, we express the vector of all random components of the model by \\ $\ds \left(\{s_n\}_{n=1}^{N}\;,\;\{s_m\}_{m=1}^{K-N}\;,\;\beta_{-K}\;,\;\beta_K\;,\{s_k\}_{k=1}^{K}\;,\;K\;,\;\lambda^*\;,\;\theta,\;W\right)$.

Initially, assume $W$ to be deterministic and define \\ $\ds \psi:=\left(\{s_n\}_{n=1}^{N}\;,\;\{s_m\}_{m=1}^{K-N}\;,\;\beta_{-K}\;,\;\beta_K\;,\{s_k\}_{k=1}^{K}\;,\;K\;,\;\lambda^*\;,\;\theta\right)$. Note that, due to redundance issues, the components of the model could be specified in other ways.

We now specify the joint distribution of all the random components of the model - $\ds (\psi|W,S)$. Note that the joint posterior density we are aiming for is proportional to this. We write the density of $(\psi|W,S)$ w.r.t. the dominating measure given by the product measure $\mathbb{Q}:=\delta^K\otimes\mathbb{G}\otimes\mathbb{L}^{K}\otimes\mathbb{L}^K\otimes\delta\otimes\mathbb{L}\otimes\mathbb{L}^{d_{\theta}}$, where $\delta$ is the counting measure, $\mathbb{G}$ is the prior GP measure on $(\beta_{-K}|\beta_K)$, $\mathbb{L}^d$ is the $d$-dimensional Lebesgue measure and $d_{\theta}$ is the dimension of the parameter vector $\theta$. This choice is related to the factorisation we choose. If we let $\mathbb{P}$ be the probability measure of our full model, the density of $(\psi|W,S)$, defined as the Radon-Nikodym derivative of $\mathbb{P}$ w.r.t. $\mathbb{Q}$, is given by
\begin{eqnarray}\label{full1}
&&\pi(\psi|W,S)=\pi(\{s_n\},\{s_m\}|\{s_k\},\beta_K,W)\pi(\beta_{-K}|\beta_K,\{s_k\},\theta)\pi(\beta_K|\{s_k\},\theta)\pi(\{s_k\}|K,S) \nonumber \\
&& \textcolor[rgb]{1.00,1.00,1.00}{\pi(\psi|W,S)=} \pi(K|\lambda^*,S)\pi(\lambda^*)\pi(\theta) \nonumber \\
&=&\Phi_N(W_N\beta_N;I_N)\Phi_{K-N}(-W_M\beta_M;I_{M})\pi_{GP}(\beta_K|\theta)\left[e^{-\lambda^*\mu(S)}\left(\lambda^*\right)^K\frac{1}{K!}\right]\pi(\lambda^*)\pi(\theta)
\end{eqnarray}
where $\{s_n\}=\{s_n\}_{n=1}^{N}$, $\{s_m\}=\{s_m\}_{m=1}^{K-N}$, $\{s_k\}=\{s_k\}_{k=1}^{K}$. $\Phi_k(\cdot;I_k)$ is the distribution function of the $k$-dimensional Gaussian distribution with mean vector zero and covariance matrix $I_k$ (k-dimensional identity matrix) and $\beta_N=(\beta_0(s_1)\ldots\beta_0(s_N)\ldots\beta_q(s_1)\ldots\beta_q(s_N))'$. Also, $W_N=(I_N\;W_1\;\ldots W_q)$, where $W_i$ is a $N\times N$ diagonal matrix with the $(n,n)$-entry being $W_i(s_n)$ - the $i$-th covariate at location $s_n$. Furthermore, $\pi_{GP}(\beta_K|\theta)$ is the density of the multivariate Gaussian process at locations $\{s_k\}_{k=1}^{K}$ w.r.t. $\mathbb{L}^K$. Furthermore, $\pi(\lambda^*)$ and $\pi(\theta)$ are the prior densities of $\lambda^*$ and $\theta$, respectively, w.r.t. $\mathbb{L}$ and $\mathbb{L}^{d_{\theta}}$. Finally, note that the joint density of the marginal model, which integrates out $\beta_{-K}$, is obtained by removing the term $\pi(\beta_{-K}|\beta_K,\{s_k\},\theta)$ after the first equality in (\ref{full1}) and is, therefore, equal to the right hand side in (\ref{full1}).

\subsection{Estimation of the intensity function}\label{sececif}

The MCMC algorithm to be proposed in Section \ref{secGS} outputs samples from the posterior distribution of the intensity function $\lambda_{S}$ at the observed locations $\{s_n\}_{n=1}^{N}$ and at another finite collection of locations which varies among the iterations of the algorithm. Nevertheless, we need to have posterior estimates of $\lambda_{S}$ over the whole space $S$. It is quite simple, though possibly costly, to sample exactly from this posterior (at any finite collection of locations). It can be done by adding an extra step to the MCMC algorithm or by a sampling procedure after the MCMC runs. Both schemes may suffer from high computational cost but the former is considerably cheaper if well-designed. Details are provided in Section \ref{efif}. Finally, efficient solutions based on lower dimension or a nearest neighbor approximation may be employed when $K$ is too large - see Section 4 of the Appendix.

\subsection{Model identifiability and practical implementation}\label{identsec}

The proposed model may suffer from identifiability problems concerning parameter $\lambda^*$. The natural way to identify it is to have this parameter as the supremum of the intensity function which, under the Bayesian approach, should be achieved by an appropriate specification of the prior distribution. Any prior that identifies this parameter to be equal or greater than the supremum solves the identifiability problem. But the former makes the parameter interpretable and optimises the computation cost (stochastically minimises $K$).

A reasonable choice for the prior of $\lambda^*$ is a (truncated) Gamma distribution for which the hyperparameter could be specified through an empirical analysis of the data set. More specifically, by obtaining an empirical estimate of the intensity in a small area with the highest concentration of points. This area should be reasonably chosen to give a good idea of the supremum of the IF. Note that the data is being used only to identify the model, which is different from using the data twice in a model which is already identified.

The prior distribution of $\beta$ may also help in the identification of $\lambda^*$. Note that, if $\beta$ is estimated to be high ($>2$) at any location, it implies that $\lambda^*$ is (practically) identified as the sumpremum of the IF. In this sense, the prior on $\beta$ may be specified in a way to favor such a scenario by, for example, fixing a positive mean parameter and/or a variance parameter coherent with the standard Gaussian distribution.

Generally speaking, identifiability is an important issue when estimating the intensity function of a non-homogeneous PP. It is well known that the reliability of the estimates relies on the amount of data available. In this sense, the higher is the actual IF the better. In a Bayesian framework, in particular, the prior on the intensity function plays an important role on the identification and estimation of this function. This is related to the fact that the data does not contain much information about the hyperparameters of GP priors. The simulated examples in Section 1 of the Appendix explore this issue and provides some insight on how to proceed in general.

Another important issue is the computational cost from dealing with GP's.
Despite their great flexibility on a variety of statistical modelling problems, Gaussian processes have a considerable practical limitation when it comes to computational cost. More specifically, simulating a $n$-dimensional GP has a cost which is typically on the order of $n^3$. This means that in our case the cost would be $O(K^3)$, without involving the procedures in Section \ref{sececif}.

Nevertheless, this issue is mitigated by a number of reasons. Firstly, our MCMC has very good properties in terms of convergence speed and autocorrelation (see Figure 3 of the Appendix) which, in turn, implies that not too many iterations are required to obtain good results. Secondly, the computational cost is feasible for reasonably sized problems due to: $i)$ matrix algebra strategies to avoid the computation of inverse matrices; $ii)$ the fast convergence of the embedded Gibbs sampling to sample from a truncated Normal; $iii)$ computational strategies to estimate the IF in a fine grid (all three points are explained in detail in Section 4 of the Appendix).

Alternative approaches based on discrete approximations are bound to suffer from similar dimensionality issues. Note however that the relevant order of magnitude there is defined by the number of grid points, which may be larger than the number of events. Also, non-trivial tuning of the MH algorithm \citep[see, for example,][]{moller} is crucial for devising an efficient MCMC algorithm to converge in feasible time. This is in sharp contrast with our algorithm that samples directly from the full conditional distribution of the GP and presents good convergence properties.

Furthermore, for the cases where the cost is still too high, some (approximating) strategies may be employed to reduce it. Most importantly, none of these defy the exactness of our methodology. Lower dimension approximations \citep[see][]{BDT,finley} may be used to deal with the GP prior and speed the computation of covariance matrices, their inverses and their Cholesky decompositions - all required for the MCMC. This type of approximation is on the second level of the model and the first level (likelihood) is still dealt with in an exact setup. \citet{simpson} (Section 3) argue that approximations in the first level have much more impact on the results. Furthermore, from a different perspective, which agrees with the argument from \citet{simpson}, the use of our proposed methodology combined with approximations for the GP may be seen as a fully exact setup where the prior on $\beta$ is given by the probability measure defined by the GP approximation. Therefore, as long as the approximation defines a Gaussian probability measure, this may be seen as an alternative model for which exact inference is carried out under our (exact) methodology. Another strategy that may reduce the computational cost is to use a carefully well designed Metropolis-type step to sample the GP. Two possible solutions are Hamiltonian MC \citep[see][]{adams} and MALA \citep[see][]{moller}. These will avoid the embedded Gibbs Sampling required to sample directly from the full conditional distribution of the GP.


\section{Computation for the spatial model}\label{seccomp}

In this Section, we present the computational details to perform inference in the spatial model.
The proposed methodology consists of a MCMC algorithm which has the exact joint posterior distribution of the unknown components of the model as its invariant distribution.
More specifically, the algorithm is a Gibbs sampling. The derivation of the full conditional distributions is not straightforward due to several reasons: intractability issues; infinite dimensionality of the parameter space; the redundancy among some of the components; the hierarchical structure of the model, specially the fact that the observations are not (explicitly) on the first level, due to thinning.
In order to sample directly from the full conditional distributions, it is essential to be able to simulate from a general class of multivariate skew-normal distributions. We define such class and propose an algorithm to sample from it.

\subsection{A general class of multivariate skew-normal distributions}\label{secSN}

We consider a general class of skew-normal distributions originally proposed in \citet{azzalini} and present it here in a particularly useful way for the context of our work. Equally important, we also propose an algorithm to sample from this distribution.

For a $d$-dimensional vector $\xi$, a $m\times d$ matrix $W$ and a $d\times d$ matrix $\Sigma$, we define
\begin{equation}\label{SNdef}
\ds                              U=\left(
                                       \begin{array}{c}
                                         U_0 \\
                                         U_1 \\
                                       \end{array}
                                  \right)
\sim N_{m+d}(0,\Sigma^*)\;\;\mbox{and}\;\;\Sigma^*=\left(
                                          \begin{array}{cc}
                                            \Gamma & \Delta' \\
                                            \Delta & \Sigma \\
                                          \end{array}
                                        \right),
\end{equation}
where $\Gamma=I_m+W\Sigma W'$ and $\Delta'=W\Sigma$. Let $a=(a_1,\ldots,a_r)>b=(b_1,\ldots,b_r)$ mean that $a_i>b_i,\;\forall i$ and define $\gamma=W\xi$. We say that $\ds (U_1+\xi|U_0>-\gamma)$ has a $SN(\xi,\Sigma,W)$ distribution whose density is given in the following proposition.

\begin{proposition}\label{SNdenslemma}
The density of $\ds (U_1+\xi|U_0>-\gamma)$ is given by
\begin{equation}\label{SNdens}
\ds f(z)=\frac{1}{\Phi_m(\gamma;\Gamma)}\phi_d(z-\xi;\Sigma)\Phi_m(Wz;I_m),
\end{equation}
where $\phi_d(\cdot;\Omega)$ and $\Phi_d(\cdot;\Omega)$ are the density and distribution function, respectively, of the $d$-dimensional Gaussian distribution with mean vector zero and covariance matrix $\Omega$.
\end{proposition}
\begin{thma}
See Appendix - Section 2.
\end{thma}

We propose the following algorithm to sample from the density in (\ref{SNdens}). Define $U_{0}^*=A^{-1}U_0$, where $A$ is the lower diagonal matrix obtained from the Cholesky decomposition of $\Gamma$, i.e. $\Gamma=AA'$. This implies that $U_{0}^*\sim N_m(0,I_m)$ and $U_0=AU_{0}^*$. We use the following results to construct our algorithm.
\begin{equation}\label{SNdensalg}
\ds f(U_1,U_0|U_0>-\gamma)=f(U_1|U_0,U_0>-\gamma)f(U_0|U_0>-\gamma).
\end{equation}

\begin{proposition}\label{SNalg1}
$\ds (AU_{0}^*|U_{0}^*\in B)$ has the same distribution as $(U_{0}|U_{0}>-\gamma)$, where $B=\{u_{0}^*\;:\;Au_{0}^*>-\gamma\}$.
\end{proposition}
\begin{thma}
See Appendix - Section 2.
\end{thma}

The decomposition in (\ref{SNdensalg}) suggests that simulation from (\ref{SNdens}) may be performed by firstly simulating $(U_0|U_0>-\gamma)$ and then using this value to simulate from $(U_1|U_0)$. Moreover, the simulation of $U_0$ is more efficient (as described in Section 3.1 of the Appendix) if we first simulate $U_{0}^*$ and then apply the appropriate transformation, as suggested by Proposition \ref{SNalg1}. The algorithm to simulate from (\ref{SNdens}) is the following.\\
\\{\scriptsize
\begin{tabular}[!]{|l|}
\hline\\
\parbox[!]{13cm}{
\texttt{
{\bf Algorithm 4.1}
\begin{enumerate}
\setlength\itemsep{-0.5em}
\item simulate a value $u^*$ from $(U_{0}^*|U_{0}^*\in B)$;
\item obtain $u=Au^*$;
\item simulate a value $z^*$ from $(U_1|U_0=u)\sim\mathcal{N}(\Delta\Gamma^{-1}u,\Sigma-\Delta\Gamma^{-1}\Delta')$;
\item obtain $z=z^*+\xi$;
\item output $z$;
\end{enumerate}
}}\\  \hline
\end{tabular}}\\
\\
The simulation of step 3 is trivial. Step 1 consists of the simulation of a truncated (by linear constraints) multivariate Normal and cannot be performed directly. The simulation from this distribution is described in the Appendix - Section 3.1.

\subsection{The Gibbs sampling algorithm}\label{secGS}

Notice that, given the data $\{s_k\}$, the remaining unknown quantities are $\left(\{s_m\},\beta_K,\beta_{-K},K,\lambda^*,\theta\right)$. We block these quantities as: $\ds\left(\{s_m\},K\right)\;\;,\;\;\beta_K\;\;,\;\;\beta_{-K}\;\;,\;\;\lambda^*\;\;,\;\;\theta$. However, blocks $\ds\beta_K$, $\lambda^*$ and $\theta$ are sampled via the collapsed Gibbs Sampling \citep{liuCGS} that integrates $\beta_{-K}$ out. This means that they are sampled from the respective full conditional distributions induced by the marginal model. We highlight the following full conditional densities - all proportional to $\pi(\psi|W,S)$ under the marginal model, accordingly.
\begin{equation}\label{full3}
\ds \pi(\beta_K|\cdot)\propto\Phi_N(W_N\beta_N;I_N)\Phi_{K-N}(-W_M\beta_M;I_{K-N})\pi_{GP}(\beta_K|\theta),
\end{equation}
\begin{equation}\label{full4}
\ds \pi(\lambda^*|\cdot)\propto\left[e^{-\lambda^*\mu(S)}\left(\lambda^*\right)^K\right]\pi(\lambda^*),
\end{equation}
\begin{equation}\label{full5}
\ds \pi(\theta|\cdot)\propto\pi_{GP}(\beta_K|\theta) \pi(\theta).
\end{equation}
The three densities above are written w.r.t. dominating measures: $\mathbb{L}^K$, $\mathbb{L}$ and $\mathbb{L}^{d_{\theta}}$, respectively, in accordance with the dominating measure used in (\ref{full1}).

Block $\beta_{-K}$ is infinite-dimensional and, therefore, is sampled retrospectively \citep[see][]{retrosp}. This means that it is sampled at a finite number of locations as these are needed in the sampling step of the first block. The term retrospective means that this sampling is actually performed when sampling the first block. This does not defy the exactness of the methodology because: i) $\beta_{-K}$ only needs to be unveiled at a finite number of locations in order to perform the other steps of the algorithm and these values are sampled exactly from the full conditional distribution of this block. Finally note that, from (\ref{full1}), this full conditional distribution is the prior Gaussian process conditional on $\beta_{K}$.

The first block $\left(\{s_m\},K\right)$ consists of the thinned events, since we are conditioning on $\{s_n\}$. We recall that, from standard results of the Poisson thinning, given the intensity function $\{\lambda(s),\;s\in S\}$, the thinned events and the observations are two independent Poisson processes with IF $\lambda^*\Phi(-W(s)\beta(s))$ and $\lambda^*\Phi(W(s)\beta(s))$, respectively. This implies directly that the full conditional distribution of the first block is a $PP(\lambda^*\Phi(-W(s)\beta(s)))$, which can be simulated by thinning events from a $PP(\lambda^*)$ with probabilities $\Phi(-W(s)\beta(s))$. Let $\beta_K$, $\theta$ and $\lambda^*$ be the current state of these component in a given iteration of the Markov chain, a sample from $\left(\{s_m\},K\right)$ is obtained by applying the following algorithm:
\\
\\{\scriptsize
\begin{tabular}[!]{|l|}
\hline\\
\parbox[!]{13cm}{
\texttt{
{\bf Algorithm 4.2}
\begin{enumerate}
\setlength\itemsep{-0.5em}
\item simulate a $PP(\lambda^*)$ on $S$ - say $\{r_1,\ldots,r_{K^*}\}$;
\item simulate $\beta_{-K}$ (retrospectively) at locations $\{r_1,\ldots,r_{K^*}\}$ from $\pi_{GP}(\cdot|\beta_K,\theta)$;
\item perform a thinning with probabilities $\Phi(-W(r_k)\beta(r_k))$, for $k=1,\ldots,K^*$;
\item output the locations retained on 3.
\end{enumerate}
}}\\  \hline
\end{tabular}}\\

\begin{lemma}\label{Alg2}
The output of Algorithm 4.2 is an exact draw from the full conditional distribution of $\left(\{s_m\},K\right)$.
\end{lemma}
\begin{thma}
See Section 2 of the Appendix.
\end{thma}

The choice of the Gaussian c.d.f. and the linearity in $\beta$ in (\ref{gpif2}) is justified by the fact that it makes it possible to sample directly from the full conditional distribution in (\ref{full3}). This leads to an algorithm with a reasonable computational cost and good convergence properties. The algorithm is the following.\\
\\{\scriptsize
\begin{tabular}[!]{|l|}
\hline\\
\parbox[!]{13cm}{
\texttt{
{\bf Algorithm 4.3}
\begin{enumerate}
\setlength\itemsep{-0.5em}
\item obtain $W_K$ from $(W_N,W_M)$ such that\\ $\Phi_K(W_K\beta_K;I_K)=\Phi_N(W_N\beta_N;I_N)\Phi_{K-N}(-W_M\beta_M;I_{K-N})$;
\item sample $\beta_K\sim SN(\mu_K,\Sigma_K,W_K)$ using Algorithm 4.1, where $\mu_K$ and $\Sigma_K$ are the mean vector and covariance matrix, respectively, of $\pi_{GP}(\beta_K|\theta)$;
\item output $\beta_K$.
\end{enumerate}
}}\\  \hline
\end{tabular}}\\

\begin{lemma}\label{Alg3}
The output of Algorithm 4.3 is an exact draw from the full conditional distribution in (\ref{full3}).
\end{lemma}
\begin{thma}
Simply note that (\ref{full3}) is proportional to the density of a $SN(\mu_K,\Sigma_K,W_K)$.
\end{thma}

The blocking and sampling schemes of our Gibbs sampling result in good convergence properties. In fact, the examples presented here suggest that convergence is attained after a few iterations.

The next step of the Gibbs sampler draws $\lambda^*$ from its full conditional distribution. This can be obtained by routine calculations: if a conjugated Gamma prior $\mathcal{G}(\alpha_{\lambda},\beta_{\lambda})$ is adopted for $\lambda^*$, its full conditional is $\mathcal{G}(\alpha_{\lambda}+K,\beta_{\lambda}+\mu(S))$.

The forth and last step from the Gibbs sampler draws $\theta$ from its full conditional distribution. This task may be carried out ordinarily - using direct simulation when possible or via an appropriately tuned MH step. There is also the option of breaking $\theta$ into smaller blocks if that is convenient for computational reasons. One attractive option is to use an adaptive Gaussian random walk Metropolis-Hastings step where the covariance matrix of the proposal is based on the empirical covariance matrix of the previous steps, as proposed by \citet{robrosARWMH}.

\subsection{Estimating functionals of the intensity function}\label{efif}

One of the purposes of fitting a Cox process to an observed point pattern is to estimate functionals of the intensity function. These functionals may include the intensity itself, the mean number of points at some subregion, etc. The estimation is performed by sampling such functionals from their posterior distribution to obtain MC estimates. This means that, on each iteration of the MCMC, this functional is sampled by computing its value given the sampled value of $\beta$. Different functions may require unveiling the value of $\beta$ at different locations.

For examples, to obtain estimates of the intensity function in the finite subset $S_0$ (a fine squared grid, say) of $S$, we need posterior samples of $(\lambda^*,\beta_{S_0})$. That is achieved by sampling $\beta_{S_0}$ from the full conditional of $\beta_{-K}$ on each iteration of the MCMC.

Another interesting functional to be estimated is the integrated intensity $\ds \Lambda(R)=\int_{R}\lambda(s)ds$ for some region $R\subseteq S$. This is the mean number of points in $R$. Monte Carlo estimates of $E[\Lambda(R)|y]$ may be obtained without any discretisation error by introducing a r.v. $U\sim\mathcal{U}(R)$ and noting that $\ds
E_U[\lambda(U)]=\frac{1}{\mu(R)}\int_R\lambda(s)ds$ \citep[see][]{bpr06a}, thus suggesting the estimator
\begin{equation}\label{MCinte}
\hat{\Lambda}(R)=\mu(R)\;\frac{1}{J}\;\sum_{j=1}^J\lambda^{(j)}(U^{(j)}),
\end{equation}
which is a strongly consistent estimator of $E[I|y]$ by the SLLN for Markov chains. A sample of $\lambda(U)$ is obtained by sampling $U$ and $\beta(U)$ on each iteration of the MCMC. The accuracy of the estimator may be improved defining a partition of $R$ and using one uniform for each subregion of the partition.

\section{Spatiotemporal model}\label{stmsec}

It is straightforward to generalise the MCMC algorithm from Section \ref{secGS} to the spatiotemporal case. We remind that $(X_0,\ldots,X_{T})$ are conditionally mutually independent homogeneous PP's on $S$, given $\lambda_\mathcal{T}^*$. The temporal dependence of the model is defined through $\beta$ (and, possibly, $\lambda_\mathcal{T}^*$).

We now write the density of $(\psi|W,S)$ with respect to the same dominating measure used in (\ref{full1}), adapted for different times, and get
\begin{eqnarray}\label{full6}
\ds \pi(\psi|W,S)&=& \prod_{t=0}^{T}\left[\Phi_{N_t}(W_{N_t}\beta_{N_t};I_{N_t})\Phi_{K_t-N_t}(-W_{M_t}\beta_{M_t};I_{K_t-N_t})\right]\pi_{GP}(\beta_{K_\mathcal{T}}|\theta) \nonumber \\
&&\times \prod_{t=0}^{T}\left[\frac{e^{-\lambda_{t}^*\mu(S)}\left(\lambda_{t}^*\right)^{K_t}}{K_t!}\pi(\lambda_{t}^*)\right]\pi(\theta)
\end{eqnarray}
where the new notation has a natural interpretation and $\pi_{GP}$ is the density of the dynamic GP in (\ref{DGP}).

We have at least two options for the blocking scheme. The first one samples\\ $\bigg(K_t,\{s_{t,m}\}\bigg)$ and $\beta_{K_t}$ separately, for each time. This algorithm may, however, lead to a chain with poor mixing properties if $T$ is large due to the temporal dependence of $\beta$ \citep[see][]{carter,fruh,danibmtk}. This problem is mitigated by a blocking scheme that makes $\bigg\{\bigg(K_t,\{s_{t,m}\}\bigg)\bigg\}_{t=0}^T$ and $\bigg\{\beta_{K_t}\bigg\}_{t=0}^T$ one block each. This choice eliminates the mixing problem mentioned above and is particularly appealing in the DGP context.

The first block is sampled by applying Algorithm 4.2 to each time $t$ and considering the GP temporal dependence on step 2. The validity of this algorithm is guaranteed by the conditional independence of the $PP(-W_t\beta_t)$'s among different times.

The full conditional density of $\beta_K$ is given by
\begin{equation}\label{full7}
\ds \pi\left(\bigg\{\beta_{K_t}\bigg\}_{t=0}^T|\cdot\right)\propto
\left[\Phi_{K_{+}}(W_{K_{+}}\beta_{K_{+}};I_{K_{+}})\pi_{GP}(\beta_{K_{+}}|,\theta)\right].
\end{equation}
where $K_{+}=\sum_{k=1}^KK_t$, $\beta_{K_{+}}$ is the vector of all $\beta_{K_t}$'s ordered in time and $W_{K_{+}}$ is the $K_{+}\times K_{+}$ matrix obtained to produce $\ds \prod_{t=0}^T\Phi_{N_t}(W_{N_t}\beta_{N_t};I_{N_t})\Phi_{K_t-N_t}(-W_{M_t}\beta_{M_t};I_{K_t-N_t})$.
The density in (\ref{full7}) implies that the full conditional distribution of $\left(\bigg\{\beta_{K_t}\bigg\}_{t=0}^T\right)$ falls into the class of skew normal distributions presented in Section 4.1. In the cases $K_{+}$ is too large, we suggest a combination of the following: i) use approximations such as the NNGP \citep{nngp} ; ii) sample $\beta_K$ in blocks of times to achieve reasonable dimension.

Block $\beta_{-K}$ is retrospectively sampled from the prior GP conditional on $\beta_K$.
The full conditional distribution of $\theta$ is carried out as before and particular blocking schemes may be motivated by the spatiotemporal structure. Finally, for a prior $\mathcal{G}(\alpha_{\lambda_t},\beta_{\lambda_t})$, the full conditional of each $\lambda_{t}^*$ is $\mathcal{G}(\alpha_{\lambda_t}+K_t,\beta_{\lambda_t}+\mu(S))$. In the case $\lambda_{t}^*=\lambda^*$, $\forall t$, the full conditional of this parameter is $\mathcal{G}(\alpha_{\lambda_t}+K_{+},\beta_{\lambda_t}+T\mu(S))$.

Extensions of the spatiotemporal model above can be proposed by adding a temporal dependence structure to $\lambda_{0:T}^*$ - this is particularly useful for prediction. One interesting possibility is the Markov structure proposed by \citet{dani3} (in a state-space model context) where $\ds \lambda_{0}^*\sim\mathcal{G}(a_0,b_0)$, $\ds \lambda_{t}^*|K_{1:t-1},\lambda_{t-1}^*=w^{-1}\lambda_{t-1}^*\varsigma_t$ and $\varsigma_t\sim Beta(wa_t,(1-w)a_t)$.
The full conditional distribution of $\lambda_{0:T}^*$ is available in Section 3.2 of the Appendix.

\subsection{Prediction}\label{secpred}

Suppose that we want to predict $Y$ at future times $\mathcal{T}^*=(T+1,\ldots,T+J)$.
The algorithm to sample from the predictive distribution of $Y_{\mathcal{T}^*}$, proceeds iteratively on time from $T+1$ onwards. Firstly, we sample $\lambda_{t}^*$ (which depends on the structure that has been adopted), then apply Algorithm 1 with the Gaussian process being simulated from $\pi_{GP}(\beta_{K_t}|\beta_{K_{\mathcal{T}}},\beta_{K_{T+1:t-1}},\theta)$. This algorithm is supported by the following result:

\begin{eqnarray}\label{a42re}
\ds &&\pi(y_{\mathcal{T}^*}|y_{\mathcal{T}})\propto\int\pi(y_{\mathcal{T}^*},\lambda_{\mathcal{T}^*}^*,\beta_{\mathcal{T}^*},\lambda_{T}^*,\beta_{\mathcal{T}},\theta|y_{\mathcal{T}})d\lambda_{\mathcal{T}^*}^*d\beta_{\mathcal{T}^*}d\lambda_{T}^*d\beta_{\mathcal{T}}d\theta \\
&&=\int\prod_{t=T+1}^{T+J}\left[\pi(y_t|\lambda_{t}^*,\beta_t)\pi(\lambda_{t}^*|\lambda_{t-1}^*,y_{t-1})\pi(\beta_{t}|\beta_{t-1},\theta)\right]\pi(\lambda_{T}^*,\beta_{\mathcal{T}},\theta|y_{\mathcal{T}})d\lambda_{\mathcal{T}^*}^*d\beta_{\mathcal{T}^*}d\lambda_{T}^*d\beta_{\mathcal{T}}d\theta, \nonumber
\end{eqnarray}
where $y_{\mathcal{T}}$ are the observed data at times $\mathcal{T}$, $\beta_{\mathcal{T}}$ is the GP at the locations of $y_{\mathcal{T}}$ and $(y_{\mathcal{T}^*},\beta_{\mathcal{T}^*})$ represent these components at a finite collection of locations at times $\mathcal{T}^*$.

The predictive distribution may be explored in different ways, specially in a point process context, by choosing convenient functions of the observations to analyse. This issue is illustrated in a simulated example presented in Section 1.2 of the Appendix. Note that the same algorithm provides prediction of the intensity function $\lambda$.


\section{Applications}\label{secsim}

We apply the proposed methodology to real and synthetic data sets to investigate its performance.
Three examples with synthetic data consider uni and bidimensional spatial models and a spatiotemporal model, respectively. Detailed results and comments can be found in Section 1 of the Appendix. Two real data sets are used in this Section to fit a spatial and a spatiotemporal model. The spatial example is also analysed using an existing approximated methodology. All the simulations concerning our methodology are coded in Ox \citep{Ox} and run in a 3.50GHz Intel i7 processor with 16GB RAM. Results obtained for synthetic data are considerably good, specially considering that data was not generated from our models for the IF. The estimates recovered well the model components, their functionals and prediction. This was achieved by fast-converging well-mixing chains, with low autocorrelation function, for all synthetic and real applications.

\subsection{Example 1. Lansing Woods data}\label{ssecsim1}

We analyse a data set referring to the location of white oak trees in a 924ft x 924ft plot in Lansing Woods, Clinton County, Michigan, USA. These data are available in the R package {\tt spatstat} \citep{brt} and have also been analysed by some other authors \citep[see, for example,][]{brt}. It consists of the location of 448 white oak trees in $S$, a re-scaled square of size 10. We adopt a constant mean function $\mu=0$ and the covariance function given in (\ref{gecf}) with $\gamma=3/2$, $\sigma^2=4$ and $\tau^2=0.5$. We also adopt a $Gamma(1,0.1)$ prior for $\lambda^*$.

\begin{figure}[!h]
\centering
  \includegraphics[scale=0.36]{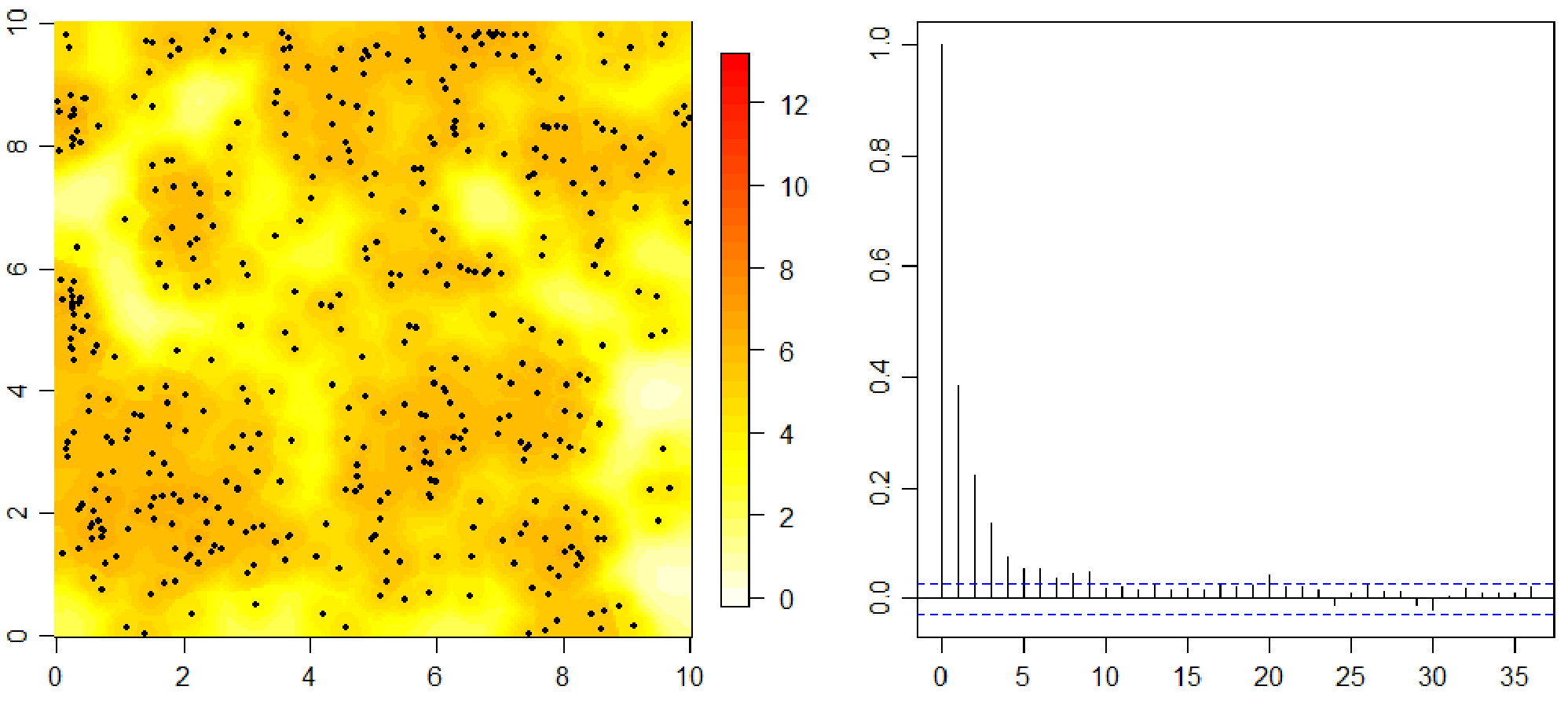}
\caption{Exact analysis. Left: map of the IF posterior mean and observed data for the White Oak example. Right: Autocorrelation function of $\Lambda([0,4]^2)$.}\label{figf6}
\end{figure}

The estimated IF obtained by our methodology is presented in Figure \ref{figf6}. It shows
a smoothly varying pattern, while also respecting the rate of occurrence of events. Results are based on a MCMC run for 5500 iterations and estimates of the IF are obtained for a burn-in of 500 iterations. Each iteration takes around 3.6s, mostly consumed in the inversion and Choleski decomposition of highly-dimensional (order $10^3$) covariance matrices and in the embedded Gibbs sampling of $\beta_K$ (3 iterations are enough). Our MCMC algorithm
allows for appropriately precise Monte Carlo estimates with a small number of iterations.
As an example, the MC estimate of the posterior mean and s.d. of $\Lambda([0,4]^2)$ are 81.8 and 6.23, respectively, with a percentage Monte Carlo error of $0.19\%$ for the mean (the number of events in this region is 93).

\begin{figure}[!h]
\centering
  \includegraphics[scale=0.32]{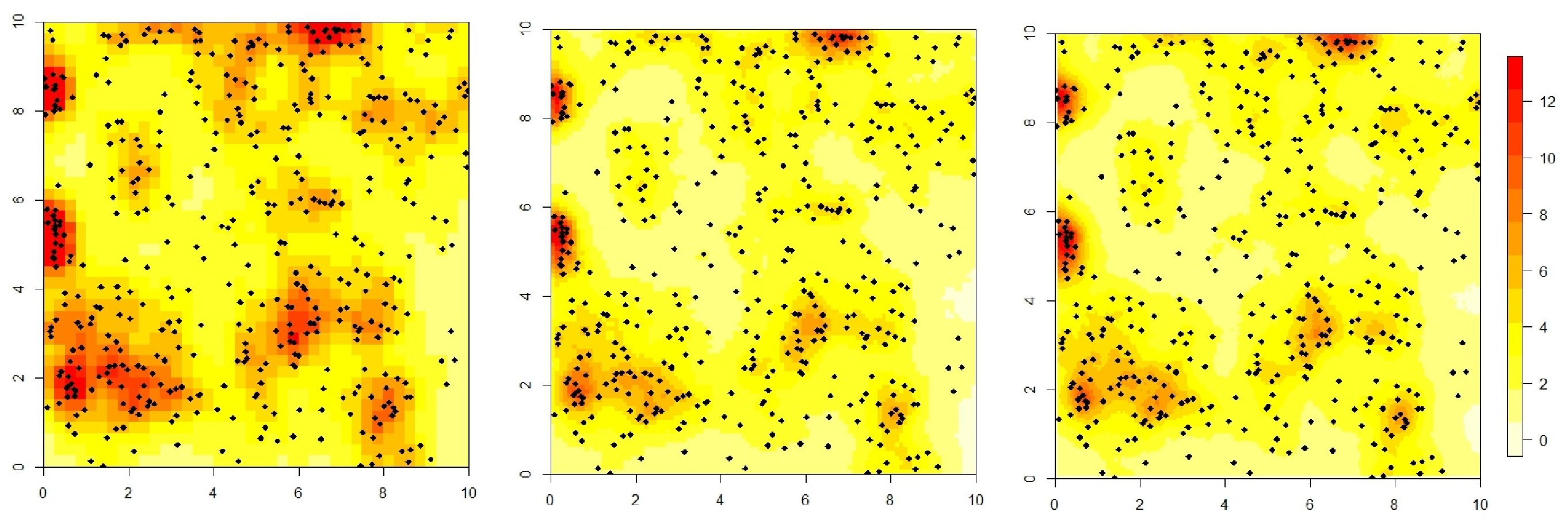}
\caption{Discretised analysis. Map of the IF posterior mean and observed data for the White Oak example, for configurations with area sizes decreasing towards 0.
Left: 1,600 areas, middle: 10,000 areas, right: 40,000 areas. All areas of each configuration are of equal size. Hyperparameters are set at $(\mu,\sigma^2,\phi)=(0,4,2.5)$ in {\tt lgcp}.}\label{figd}
\end{figure}

A discretisation-based approach for the LGCP \citep{moller} is also presented for comparative purposes. The discretised methodology was run in the R package {\tt lgcp} \citep{bendig}.
A number of hyperparameters values were considered and we report the smoothest estimate obtained. Figure \ref{figd} shows the estimate of the IF for this methodology. The results indicate convergence to an unknown limit,
qualitatively similar (but not equal) to our results. This was to be expected as the underlying models are not the same.  In contrast, our results are already the estimates of a continuously varying IF.

\subsection{Example 2. New Brunswick fires}\label{ssecsim2}

\begin{figure}[!h]
\centering
  \includegraphics[scale=0.35]{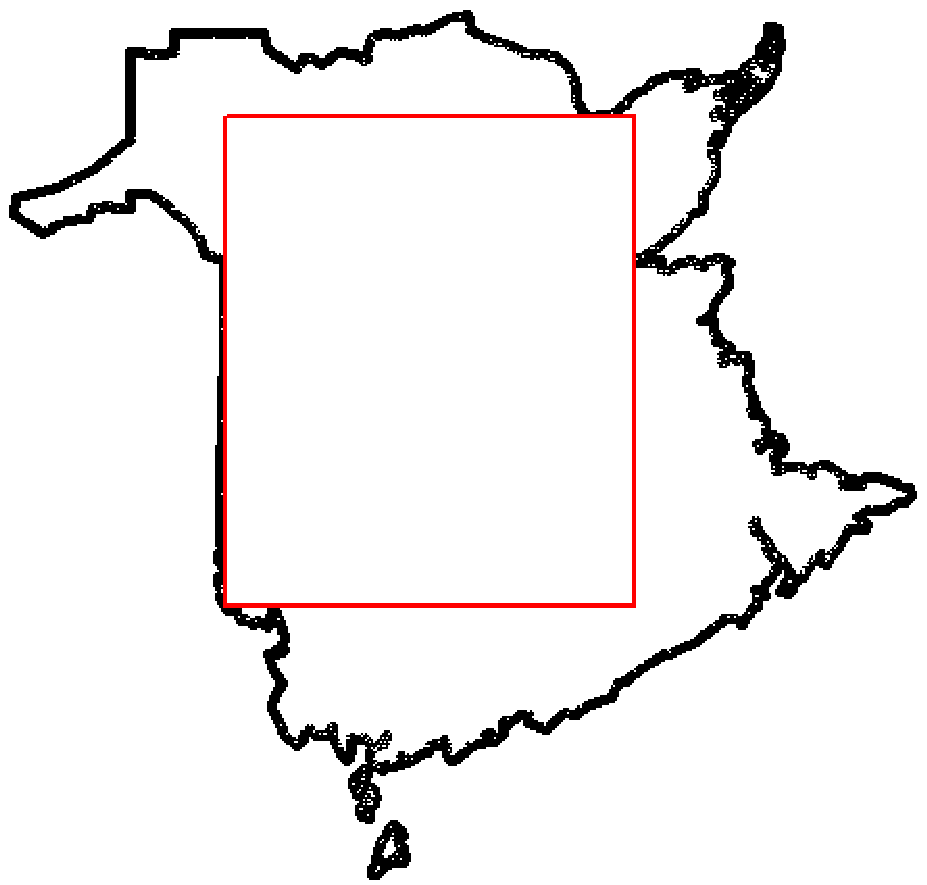}
\caption{The province of New Brunswick and the area considered in the analysis.}\label{figf8}
\end{figure}

This dataset is also provided in the R package {\tt spatstat}. It is provided by the Department of Natural Resources of the province of New Brunswick, Canada, and consists of all fires falling under their jurisdiction for the years 1987 to 2003 inclusive (with the year 1988 omitted). We consider fires notified in the rectangular area (rescaled to $54.5\times45.5$) that is shown in Figure \ref{figf8}. Table \ref{tabt6} gives the number of fires per year.

\begin{table}[!h]
 {\centering {\scriptsize
\begin{tabular}{|c|c|c|c|c|c|c|c|c|}
  \hline
  Year        & 1987 & 1989 & 1990 & 1991 & 1992 & 1993 & 1994 & 1995  \\
  N. of fires & 216  & 120  & 102  & 211  & 155  & 123  & 136  & 169   \\  \hline
  Year        & 1996 & 1997 & 1998 & 1999 & 2000 & 2001 & 2002 & 2003 \\
  N. of fires & 122  & 94   & 86   & 224  & 140  & 194  & 127  & 94 \\ \hline
\end{tabular}}
  \caption{Number of fires per year for the New Brunswick data.}\label{tabt6}}
\end{table}

\begin{figure}[!h]
\centering
  \includegraphics[scale=0.35]{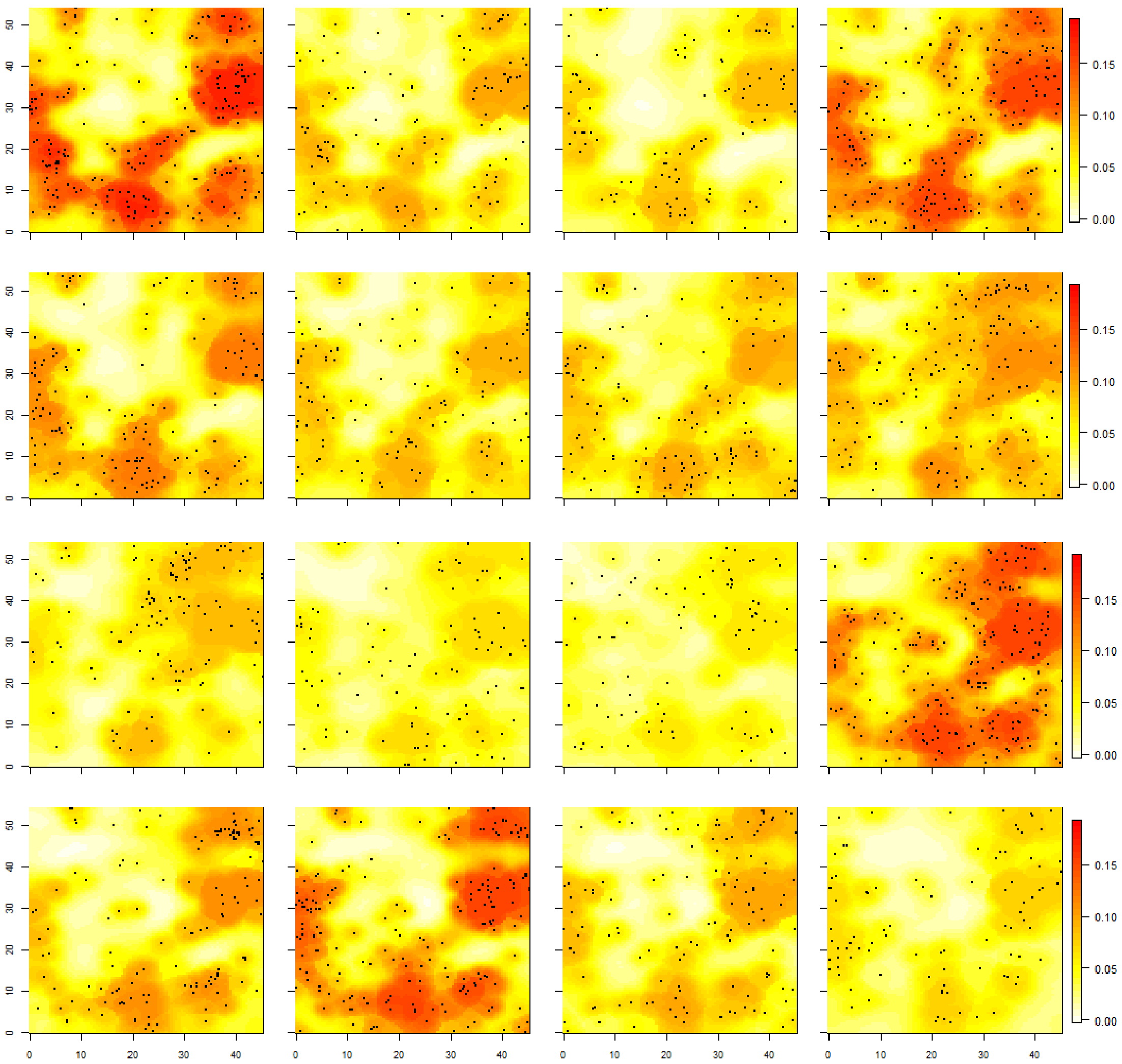}
\caption{Maps of the IF posterior mean in the New Brunswick fires example. Years are ordered from top row to bottom row, then from left to right in each row at the sequence 1987, 1989, 1990, ... , 2003.}\label{figf10}
\end{figure}

We fit the dynamic model with $\ds f(\beta_t(s),W_t(s) ) =\beta_{0,t}(s)$ and $\beta_{0,t}(s)=\beta_{0,t-1}(s) + w_{t}(s)$, where $\beta_{0,0}$ and $w_t$ are Gaussian processes with the covariance function given in (\ref{gecf}) and hyperparameters $(0,1.75^2,10)$, $(0,0.5^2,15)$, respectively. We also adopt a time-independent structure for the $\lambda_{t}^*$ parameters with $\lambda_{t}^*\sim Gamma(15,100)\mathds{1}(<0.4)$, $\forall\;t$. The estimated IF is shown in Figure \ref{figf10}. The results clearly shows not only spatial but also temporal smoothness of the IF. The panels associated with each year exhibit similarity of spatial patterns over consecutive years as a direct consequence of
the temporal dependence of our dynamic GP. This effect is more vividly seen at the edges of the area of study, in a U-shaped region with higher values of the IF, specially at the beginning and the end of the time window considered but still bearing compromise with the observed events.

\newpage


\section{Final remarks}\label{secconc}

This paper proposes a novel methodology to perform exact Bayesian inference in spatio-temporal Cox processes in which the intensity function dynamics is described by a multivariate Gaussian process.
We showed how usual components of spatio-temporal point patterns such
as trend, seasonality and covariates can be incorporated, with
flexibility of their effects warranted by the Gaussian process prior.

The methodology is exact in the sense that no discrete approximation of the process is used and Monte Carlo is the only source of inaccuracy. Inference is performed via MCMC, more specifically, a Gibbs sampler whose particular choice of blocking and sampling scheme leads to fast convergence. The validity of the methodology is established through the proofs of the main results. Finally, simulated and real data studies illustrate the methodology and provide empirical evidence of its efficiency.

This work may give rise to new problems and possibilities that may be considered in future work. An immediate extension of our models involves consideration of marks to the Poisson events. These marks may be described with a variety of components, whose effects are allowed to vary smoothly, in line with the models used for the IF. The introduction of non-spatiotemporal covariates is another extension that may be a very useful contribution to a number of areas of application.
Finally, computational developments are still required to deal with very large data sets, which is a general problem when working with Gaussian processes.
Computation with GPs is an area of very active and promising research, that is likely to bring computational gains to our methodology against other existing methodologies in the near future. As a result, we can envision the development of a software implementing our methods in the near future.

\section*{Acknowledgements}

The authors would like to thank the referees and the Associate Editor for many useful comments that led to a much more improved version of the paper throughout. They also thank Gareth Roberts and Krzysztof {\L}atuszi\'{n}ski for insightful discussions about MCMC, Piotr Zwiernik for insightful discussions about matrices, Ryan Adams for providing the Matlab code to run \citet{adams}'s algorithm, Jony A. Pinto Jr and Jesus E. Gamboa for their help on the data analyses with the discretised versions of the models and Daniel Simpson for drawing our attention to {\tt spatstat}. The first author thanks FAPEMIG for financial support and the second author thanks CNPq-Brazil and FAPERJ for financial support.

\bibliographystyle{apalike}
\bibliography{biblio}

\newpage

\setcounter{section}{0}


{\centering \bf\Huge Appendix}

\section{Simulated examples}\label{secsim}

\subsection{Simulated examples - spatial models}\label{secsim1}

The data was simulated from a Poisson process with IF $\lambda(s)=2\exp(-s/15)+\exp(-(s-25)^2/100)$, for $s\in[0,50]$. We apply the inference methodology proposed in this paper to study its efficiency and robustness under different prior specifications. We assume that $\beta$ is a Gaussian process with constant mean function $\mu$ and the covariance function given in (7) with $\gamma=3/2$.

\begin{figure}[!h]
\centering
   \includegraphics[scale=0.4]{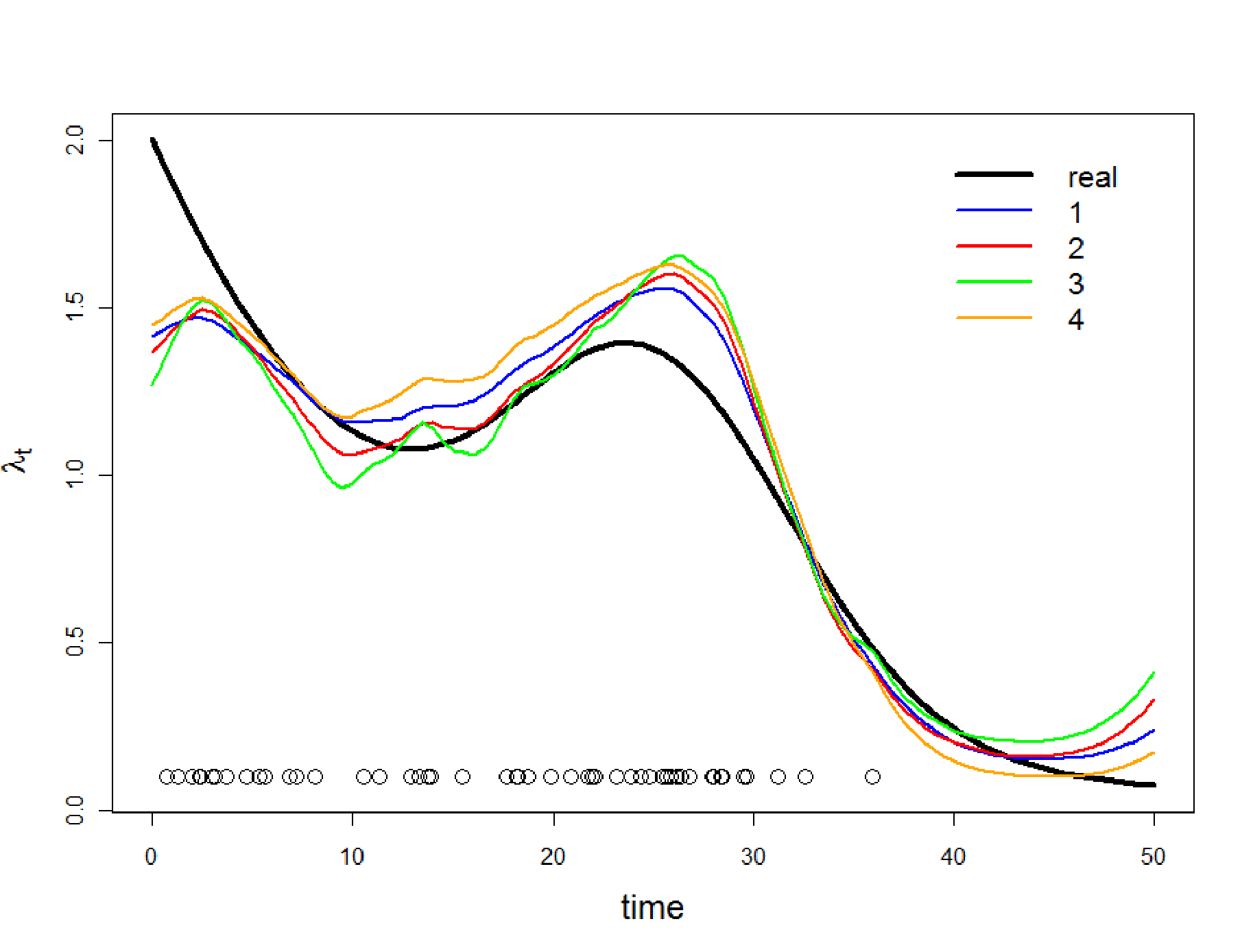}
\caption{Real and estimated (posterior mean) IF and realisation of the unidimensional Poisson process.}\label{fig2}
\end{figure}

An extensive analysis indicated that the posterior distribution of the intensity function is sensitive to the prior specification of the GP, as expected. Also, data does not contain much information about the hyperparameters of the GP and, as a consequence, non-informative priors lead to high posterior variance for those and unstable estimates of the intensity function. Efficient estimation requires the hyperparameters to be fixed or to have informative priors. Reasonable choices forthe hyperparameters' values can be obtained with some knowledge about the dynamics of Gaussian processes. In particular the chosen values should reflect the expected smoothness for the intensity function. Parameter $\lambda^*$ is also sensitive to prior specification but good results are obtained following the guidelines of Section 3.3.

We consider four different prior specifications. All of them consider a $\mathcal{G}(2.2,1.5)$ prior for $\lambda^*$. The first three specifications fix the parameter vector $\theta=(\mu,\sigma^2,\tau^2)$ at $(0,1,20)$, $(0,1,10)$ and $(0,1,5)$, respectively. The last case fixes $\mu=0$ and estimates $(\sigma^2,\tau^2)$ with uniform priors $\sigma^2\sim\mathcal{U}(0.25,4)$ and $\tau^2\sim\mathcal{U}(1,30)$. The estimated intensity function in each case is shown in Figure \ref{fig2}. Results are similar for all prior specifications which reinforces the idea that reasonable choices for the hyperparameters lead to good results. The influence of the prior is clearly seen as the estimated function is smoother for higher choices of $\tau^2$. The trace plots of $\lambda^*$ and $(\sigma^2,\tau^2)$ (in case 4) suggest fast convergence of the chain. The posterior distribution of $\lambda^*$ is very similar in all cases, with mean around 2.37 and s.d. 0.51. The parameter vector $(\sigma^2,\tau^2)$ in case 4 has large posterior variance indicating that the data does not have much information about it. The posterior mean and standard deviation of $(\sigma^2,\tau^2)$ are $(2.59 , 20.79)$ and $(0.94 , 6.84)$, respectively.

\begin{figure}[!h]
\centering
   \includegraphics[scale=0.40]{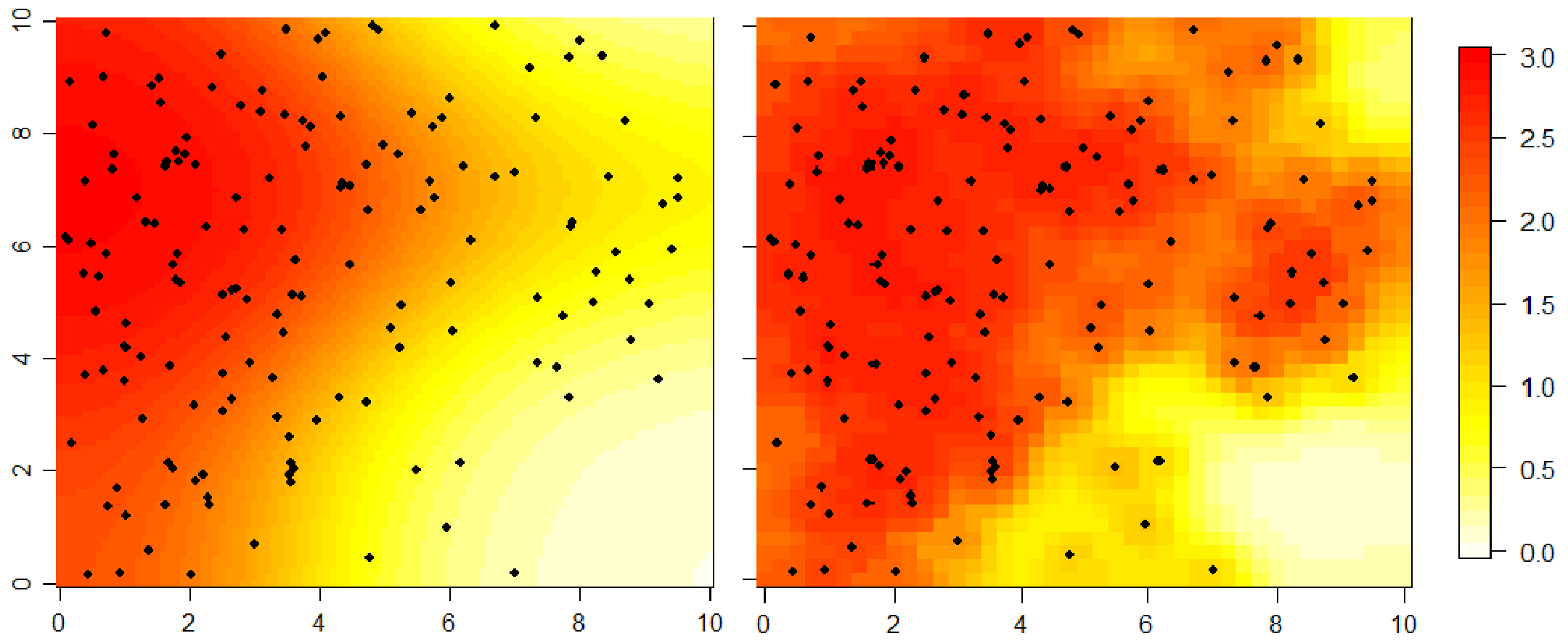}
\caption{Real (left) and estimated (right) IF and realisation of the bidimensional Poisson process.}\label{fig3}
\end{figure}

Data was also simulated from a bidimensional Poisson process on $[0,10]\times[0,10]$ with IF $\lambda(s)=3\Phi(\beta^{(0)}(s))$, where $\beta^{(0)}(s)=(8/3)\exp\{-s_{(1)}^2/30\}+(4/3)\exp\{-(s_{(2)}-7)^2/12\}-2$.
We assume the same covariance function as in the unidimensional example and set the values $(\mu,\sigma^2,\tau^2)=(0,1.8^2,1.5)$ for $\beta^{(0)}$. Figure \ref{fig3} shows real and estimated intensity function and the realisation of the process. It is clear that the intensity function is well estimated.

\begin{figure}[!h]
\centering
   \includegraphics[scale=0.45]{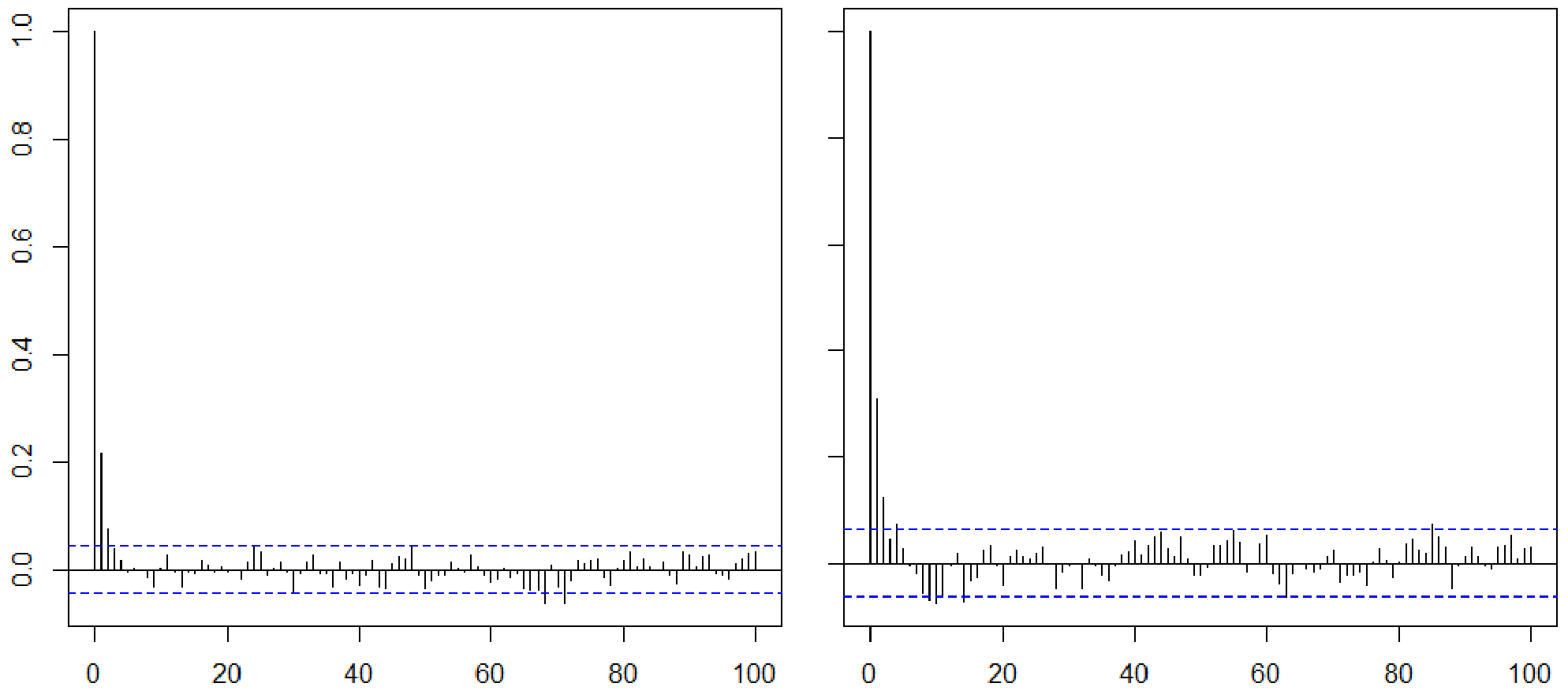}
\caption{Autocorrelation of the integrated intensity for the uni (left) and bi (right) dimensional examples.}\label{figac}
\end{figure}

Figure \ref{figac} illustrates the good convergence properties of our MCMC. For the univariate example, it refers to the chain of $\Lambda([0,50])$, for which the MC estimate of its posterior mean is 53.54 with a percentage MC error of $0.38\%$ (for a sample of size 3000). For the bidimensional example, it refers to $\Lambda([0,5]\times[0,10])$ and has MC estimate of its posterior mean equals to 121.05 with percentage MC error of $0.52\%$ (for a sample of size 1000).

The average CPU time per iteration of the MCMC for the univariate example was 0.036 seconds without estimating the IF in a grid and  0.45 seconds when estimating it in a grid of 500 locations (sampling directly from the GP predictive distribution). For the bivariate example, those values are 0.23 seconds and 1.57 seconds, respectively, for a grid of $1,600$ locations (sampling directly from the GP predictive distribution). Results suggest that convergence is rapidly attained (under 50 iterations). Also, the MCMC has very good mixing properties - see Figure \ref{figac}.

\subsection{Simulated examples - spatiotemporal model with a seasonal component}\label{secsim3}

We consider a spatiotemporal example with a seasonal component whose effect varies in space according to $\ds f(\beta_t(s),W_t(s) ) =\beta_{0,t}(s)+\beta_{1}(s)\cos\left(2\pi t/p+\phi\right)$ and $\beta_{0,t}(s)=\beta_{0,t-1}(s) + w_{t}(s)$, $\beta_{0,1}$, $\beta_1$ and $w_t$ are Gaussian processes with the covariance function given in (7) and hyperparameters $(0,4,5)$, $(0,0.7^2,10)$ and $(1,1.5^2,5)$, respectively. The data is simulated considering $(-0.2,1.8^2,15)$ and $(0,0.5^2,20)$ as GP hyperparameters for $\beta_{0,0}$ and $w_t$, respectively, and a deterministic $\beta_1=2.4\exp\{-s_{(1)}^2/25\}+0.6\exp\{-(s_{(2)}-7)^2/36\}-0.288$. Furthermore, we set $\lambda_{t}^*=1.5,\;\forall t$ and $\phi=\pi/2$ (for the generation and for the analysis).

\begin{figure}[!h]
\centering
   \includegraphics[scale=0.35]{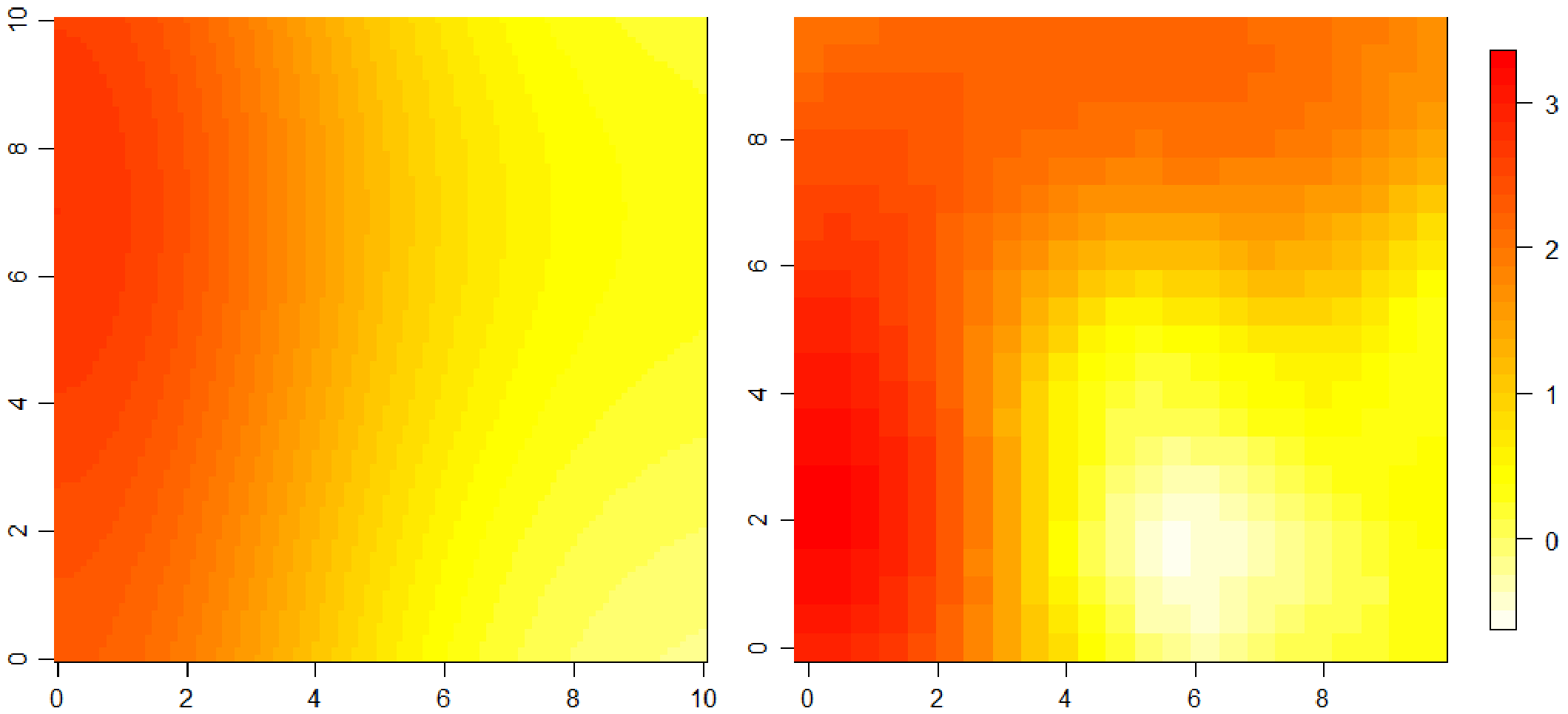}
\caption{Seasonal effect $\beta_1$. Real (left) and estimated (right).}\label{f5b}
\end{figure}

\begin{figure}[!h]
\centering
   \includegraphics[scale=0.4]{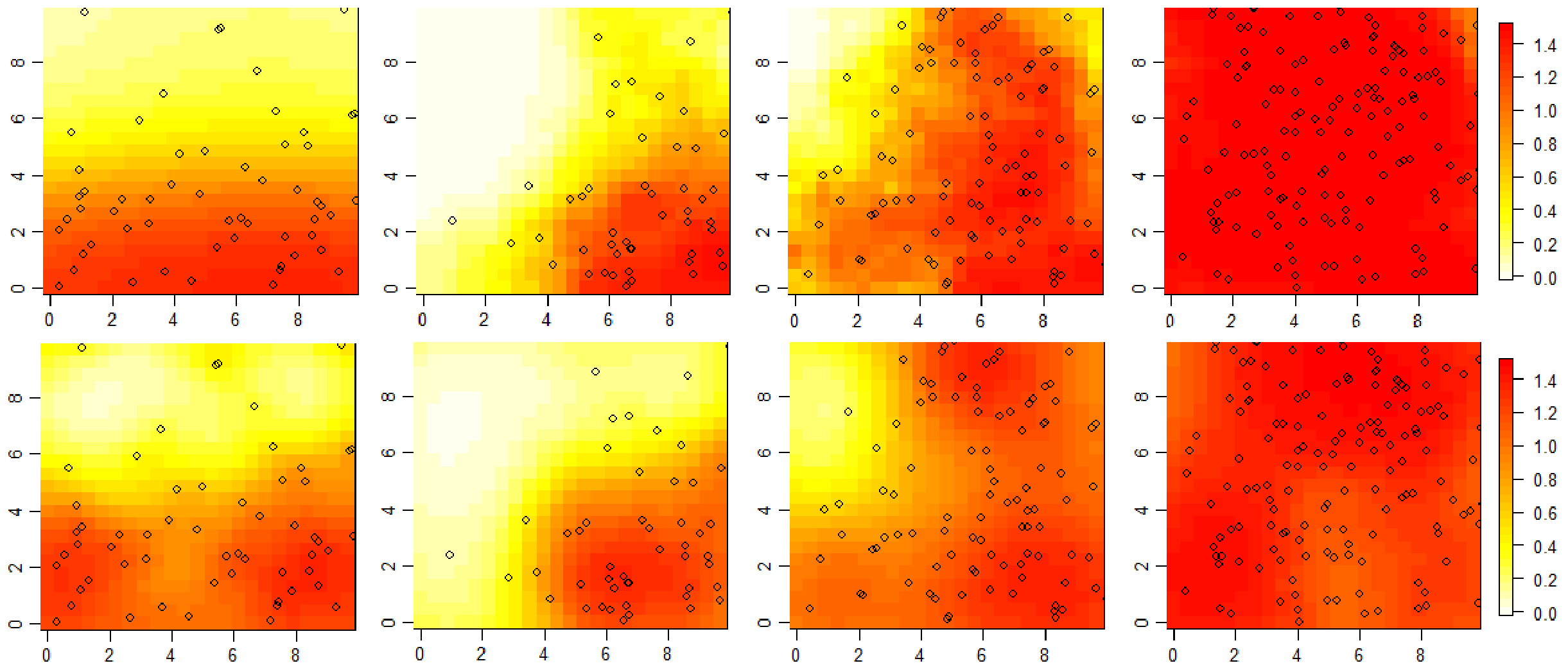}
\caption{True (top) and estimated (bottom) IF for times 0, 5, 10 and 15.}\label{f5a}
\end{figure}

\begin{figure}[!h]
\centering
   \includegraphics[scale=0.55]{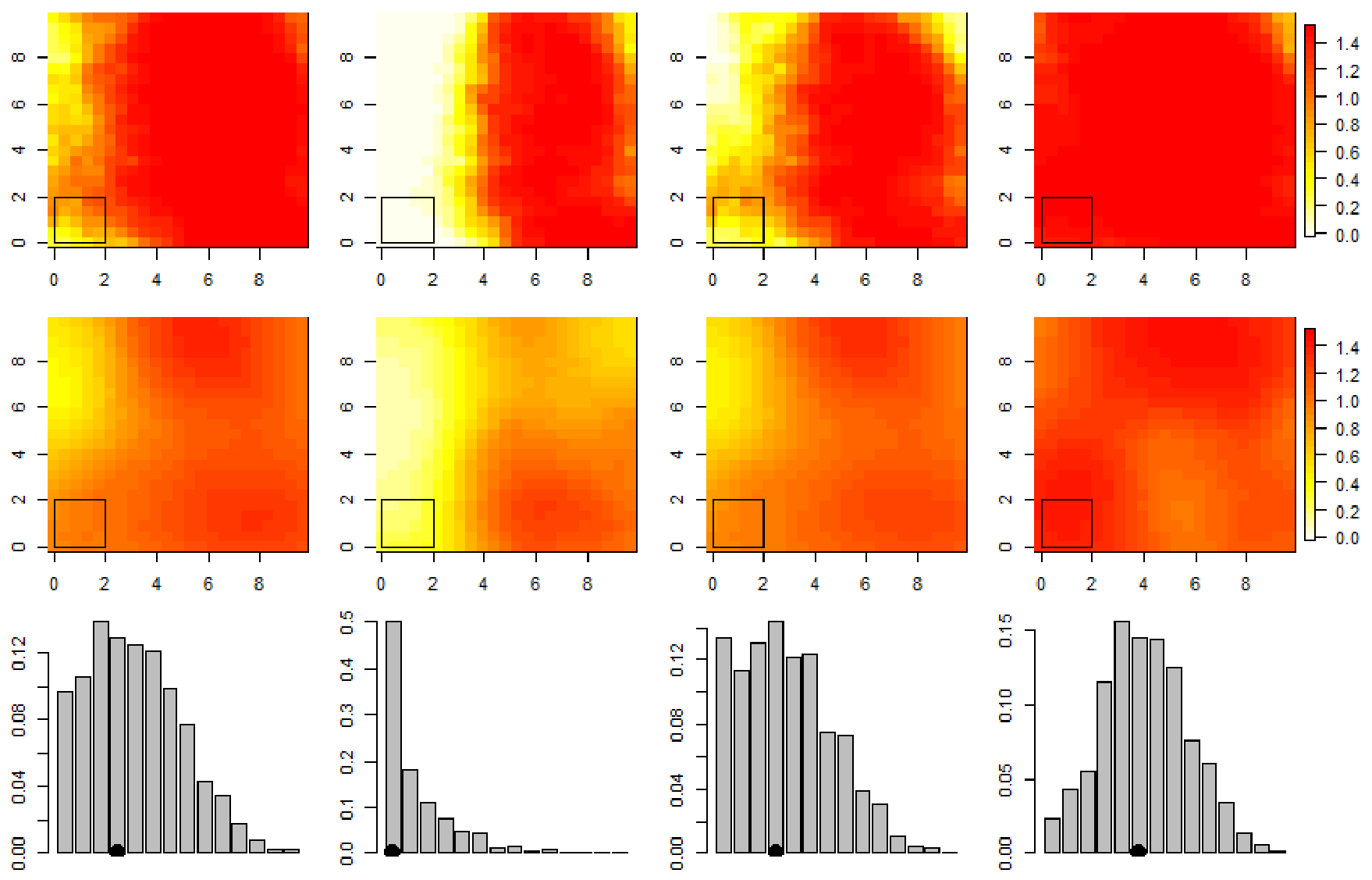}
\caption{Prediction for times 16, 17, 18 and 19. Top: true IF; middle: predictive mean of the IF; bottom: predictive distribution for the number of points in $[0,2]\times[0,2]$, the black dots represent the respective (future) true values for one replication of $Y$.}\label{f5c}
\end{figure}

Figure \ref{f5b} shows the estimation of the space-varying seasonal coefficient to be satisfactory, capturing the dip in the (south)eastern portion of the area of study. Figure \ref{f5a} shows that the IF is very well estimated at a wide selection of times. Figure \ref{f5c} shows prediction results. Prediction for the (latent) IF and for the (observed) number of points in $[0,2]\times[0,2]$ for
future time instants is also good. Finally, we also predict the average number of points in the whole space at time $T=15$ using estimator (19), which returned the value 103.45. The expected value based on the true model is 121.58.

\newpage

\section{Proofs of main results}

\subsection*{Proof of Proposition 1}

Firstly, we prove that $\Gamma$ and $\Sigma^*$ are positive definite matrices, for any $m\times d$ real matrix $W$ and positive definite $d\times d$ matrix $\Sigma$. Let $X$ and $\varepsilon$ be r.v.'s such that $X\sim N(0,\Sigma)$ and $\varepsilon\sim N(0,I_m)$. Now define a r.v. $Y=WX+\varepsilon$. This implies that $Cov(Y,Y)=I_m+W\Sigma W'=\Gamma$ and $\Gamma\succ0$. Now, the Schur complement of $\Sigma^*$ is given by
$S=\Gamma-\Delta'\Sigma^{-1}\Delta=\Gamma-W\Sigma W'=I_m$. The fact that $S\succ0$ and $\Sigma\succ0$ implies that $\Sigma^*\succ0$.

Now, by standard properties of the multivariate normal distribution we have that $(U_0|U_1=z^*)\sim N_m(\Delta'\Sigma^{-1}z^*,\Gamma-\Delta'\Sigma^{-1}\Delta)$, where $\Gamma-\Delta'\Sigma^{-1}\Delta=I_m$. Therefore, by the symmetry of the standard Gaussian cdf and the Bayes Theorem, we have that the density of $\ds (U_1|U_0>-\gamma)$ is given by
$$\ds f(z^*)=\frac{f_{U_1}(z^*)P(U_0>-\gamma|U_1=z^*)}{P(U_0>-\gamma)}=\frac{\phi(z^*,\Sigma)\Phi_m(\gamma+\Delta'\Sigma^{-1}z^*;I_m)}{\Phi_m(\gamma;\Gamma)}.$$
Finally, we apply the transformation theorem to find the density of $\ds (U_1+\xi|U_0>-\gamma)$.

\subsection*{Proof of Proposition 2}

The density $f$ of $\ds(U_{0}^*|U_{0}^*\in B)$ is given by
$$\ds f(u)=\frac{1}{P(U_{0}^*\in B)}\phi_m(u;I_m)\mathds{1}(u\in B).$$
Applying the transformation theorem to find the density $f^*$ of $(AU_{0}^*|U_{0}^*\in B)$, we get
$$\ds f^*(u)=\frac{1}{\Phi_m(\gamma;\Gamma)}\phi_m(u;\Gamma)\mathds{1}(u>-\gamma).$$

\subsection*{Proof of Lemma 1}

Conditional on the whole GP $\beta$ and $\lambda^*$, therefore, on the IF, standard properties of Poisson processes imply that the thinned events are a Poisson process with IF $\lambda^*\Phi(-W(s)\beta(s))$, independent of the data. For that reason, they are sampled by thinning the events of a $PP(\lambda^*)$.




\section{Sampling algorithms}

\subsection{Algorithm to sample from a truncated multivariate Normal distribution}

Firstly note that our aim is to simulate from a multivariate Normal distribution with independent coordinates restricted to a region $R$ defined by linear constraints, more specifically $R:=\{u_{0}^*,\;Au_{0}^*>-\gamma\}$. Moreover, note that $A$ is the lower triangular matrix obtained from applying the Cholesky decomposition to $\Gamma$.

The most obvious way to simulate exactly from a truncated multivariate Normal distribution is by using a rejection sampling algorithm that proposes from the non-truncated distribution. In this case, the global acceptance probability of this algorithm is equal to the probability of the truncated region. However, this probability is typically going to be very small in high dimensions, making this algorithm very inefficient.

A more efficient alternative is provided by MCMC, more specifically, a Gibbs sampling. This method could be applied directly to $U_0$ but the resulting chain would have much higher correlation among the blocks of the Gibbs sampler, which would considerably slow down its convergence \citep[see][]{RDavis}.

The Gibbs sampler alternates among the simulation of
$(U_{0,i}^*|U_{0,-i}^*)$, $\;i=1,\ldots,m$, where $U_{0,-i}^*=U_{0}^*\setminus\{U_{0,i}^*\}$, which are all univariate standard Gaussians restricted to $R$. Basically, for a given $i$, $R$ consists of $(m-i+1)$ linear inequalities and has the form $(\max\{l_j\},\min\{L_j\})$, where $l_j$ and $L_j$ are the lower and upper limits, respectively, of each inequality. Note that the diagonal of $A$ is strictly positive ($\Gamma$ is positive definite) which means that the lower limit will always be a real number whereas the upper limit may be $+\infty$.

In order to favor a faster convergence, we choose an initial value that already belongs to $R$. This is trivially obtained by taking advantage of the triangular form of $A$ and simulating $U_{0}^*$ recursively from $U_{0,1}^*$ onwards. The algorithm is as follows:\\
\\{\scriptsize
\begin{tabular}[!]{|l|}
\hline
\parbox[!]{13cm}{
\texttt{
\begin{enumerate}
\setlength\itemsep{-0.5em}
\item Simulate the initial value of the chain and make $k=1$;
\item for $i$ in $1:m$ do:
\begin{enumerate}[label*=\arabic*.]
\setlength\itemsep{-0.5em}
  \item simulate ${u_{0,i}^*}^{(k)}$ from $({U_{0,i}^*}^{(k)}|{U_{0,1}^*}^{(k)},\ldots,{U_{0,i-1}^*}^{(k)},{U_{0,i+1}^*}^{(k-1)},\ldots,{U_{0m}^*}^{(k-1)})$;
\end{enumerate}
\item if $k$ has reached the desired number of iterations, STOP and OUTPUT the last sampled value of $U_{0}^*$; ELSE, GOTO 2;
\end{enumerate}
}}\\  \hline
\end{tabular}}

\subsection{Results for the dynamic approach of $\lambda^*$}

We present here the algorithm to sample from the full conditional distribution of $\lambda_{1:T}^*$ when the following dynamic model is assumed:
\begin{eqnarray}
  \ds \lambda_{0}^*&\sim&\mathcal{G}(a_0,b_0) \nonumber \\
  \ds \lambda_{t}^*|K_{1:t-1},\lambda_{t-1}^*&=&w^{-1}\lambda_{t-1}^*\varsigma_t \nonumber \\
  \varsigma_t&\sim& Beta(wa_t,(1-w)a_t) \nonumber
\end{eqnarray}

The algorithm is a FFBS (Forward Filtering Backward Sampling) and is an adaptation of the results from \citet{dani3}. It proceeds as follows.\\
\\{\scriptsize
\begin{tabular}[!]{|l|}
\hline
\parbox[!]{13cm}{
\texttt{
\begin{enumerate}
\setlength\itemsep{-0.5em}
\item For $t$ in $1:T$, compute: $a_t=wa_{t-1}+K_t$, $b_t=wb_{t-1}+\mu(S)$;
\item sample $\lambda_{T}^*\sim Gamma(a_T,b_T)$;
\item for $t$ in $(T-1):0$, sample: $\lambda_{t}^*=w\lambda_{t+1}^*+L_t$, where $L_t\sim Gamma((1-w)a_t,b_t)$.
\end{enumerate}
}}\\  \hline
\end{tabular}}

\section{Strategies to improve the computation cost}

As it has been mentioned before, the cost of the proposed MCMC algorithm is $O(K^3)$. Nevertheless, some features of the algorithm and some computational strategies make this cost reasonable for considerably high values of $K$.

One of the most potentially expensive steps of the algorithm is the embedded Gibbs sampler used to sample from the truncated multivariate Gaussian distribution, which is required to sample from the full conditional distribution of $\beta_K$. Basically, its cost increases with its dimension ($K$) and the number of iterations. However, since the $A$ matrix that defines the linear constraints of the truncation is lower triangular, we are able to use an initial value for the embedded chain is which already in the truncated region. This implies that a small number of iterations is enough to get convergence, even in high dimensions. Empirical results suggest that 3 to 5 iterations are enough in all the examples presented in this paper.

Another potentially expensive step is the estimation of the IF in a fine mesh $S_0$, as described in Section 3.2. In order to reduce the computational cost, one may, for example, store the sum of $\beta_{S_0}$ over the iterations for each location and output its posterior mean. Moreover, there is no particular reason to store all the draws from $\beta_K$, which also reduces the computational cost considerably. Another option to reduce computational cost is to use the observed and thinned locations as part of the grid and add extra locations to $S_0$ as necessary. If $S_0$ is a small set, it is computationally feasible to store the whole posterior sample of $\lambda_{S_0}$ and compute, for example, credibility intervals and/or posterior marginal densities.
Furthermore, lower dimension approximations may lead to suitable choices for sampling the Gaussian process.
A simple and efficient solution is proposed in \citet{BDT}. Another strategy, which is particularly more efficient for high values of $K$, is to use a nearest neighbor approximation. This means that the value of the IF at a given location is approximated by the value at the nearest location from $\{s_k\}$, in each iteration of the MCMC.

\section{Theoretical and empirical concerns about the validity of Adams et al.'s algorithm}

The validity of the MCMC algorithm from \citet{adams} seems to be compromised by the MH step where the number and location of the thinned events are sampled.
Firstly, note that this MH step is not a standard MH algorithm. It involves a dimension-changing chain with both discrete and continuous coordinates. This means that one should consider the most general formulation of the MH algorithm that accounts for general state spaces - this is well explained in \citet{Tirney}. Basically, defining $Q(x,dy)$ as the proposal measure of the MH chain and $\pi(dy)$ as the target measure, the acceptance probability of a move from $x$ to $y$ is given by $\min\{1,r(y,x)\}$, where $r(y,x)$ is the Radon-Nikodym (RN) derivative of $\pi(dy)Q(y,dx)$ w.r.t. $\pi(dx)Q(x,dy)$. This includes the standard and most popular version of the Metropolis algorithm where each of the four individual measures are absolutely continuous w.r.t. a common dominating measure - generally Lebesgue. \citet{adams}, however, do not seem to consider this general formulation and devise the acceptance probability by writing the RN derivative of each of the four individual measures w.r.t. distinct dominating measures.

\begin{figure}[!h]
\centering
   \includegraphics[scale=0.3]{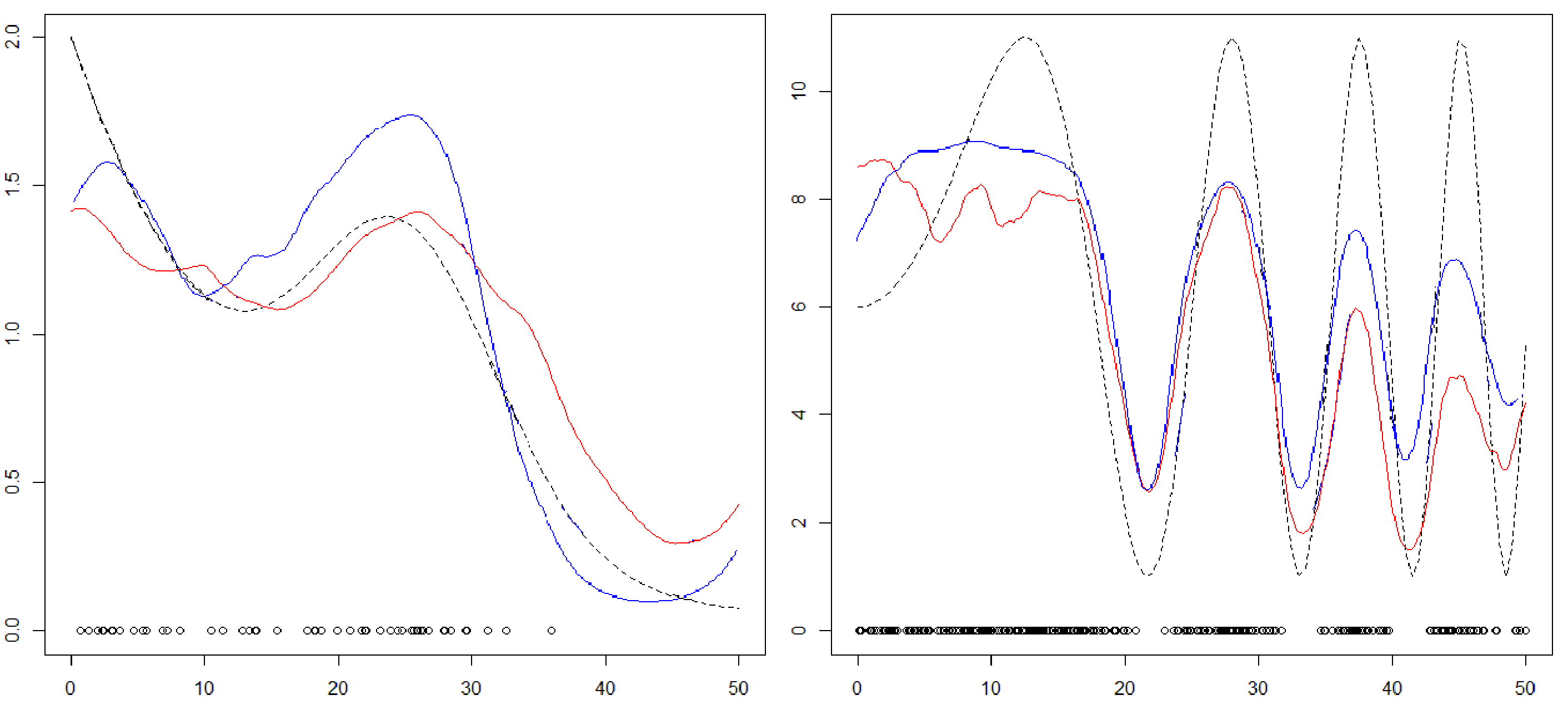}\\
\caption{Real (dashed) and posterior mean IF for ours (blue) and Adams et al.'s (red).}\label{fig4if}
\end{figure}

We compare the algorithm from Adam's et al.'s to ours in two examples, using the same GP prior with the same fixed hyperparameter values (see Figure \ref{fig4if}). The true IF's are $\lambda(s)=2\exp(-s/15)+\exp(-(s-25)^2/100)$ and $\lambda(s)=5\sin(s^2/100)+6$. The results can be compared via the integrated distance, given by $d(\lambda,\hat{\lambda})=\int_S|\lambda(s)-\hat{\lambda}(s)|ds$, where $\lambda$ is the true IF and $\hat{\lambda}$ is the respective estimate. They show a better performance of our models with $d(\lambda,\hat{\lambda})=8.371$ (ours) and
$d(\lambda,\hat{\lambda})=9.010$ (Adams et al.) for results from the first example and $d(\lambda,\hat{\lambda})=84.744$ (ours) and $d(\lambda,\hat{\lambda})=98.885$ (Adams et al.) for results from the second example. Note that, if both algorithms were correct, their estimates should be very similar - the only source of difference being in the logit/probit specification (and Monte Carlo variation).

\begin{figure}[!h]
\centering
   \includegraphics[scale=0.28]{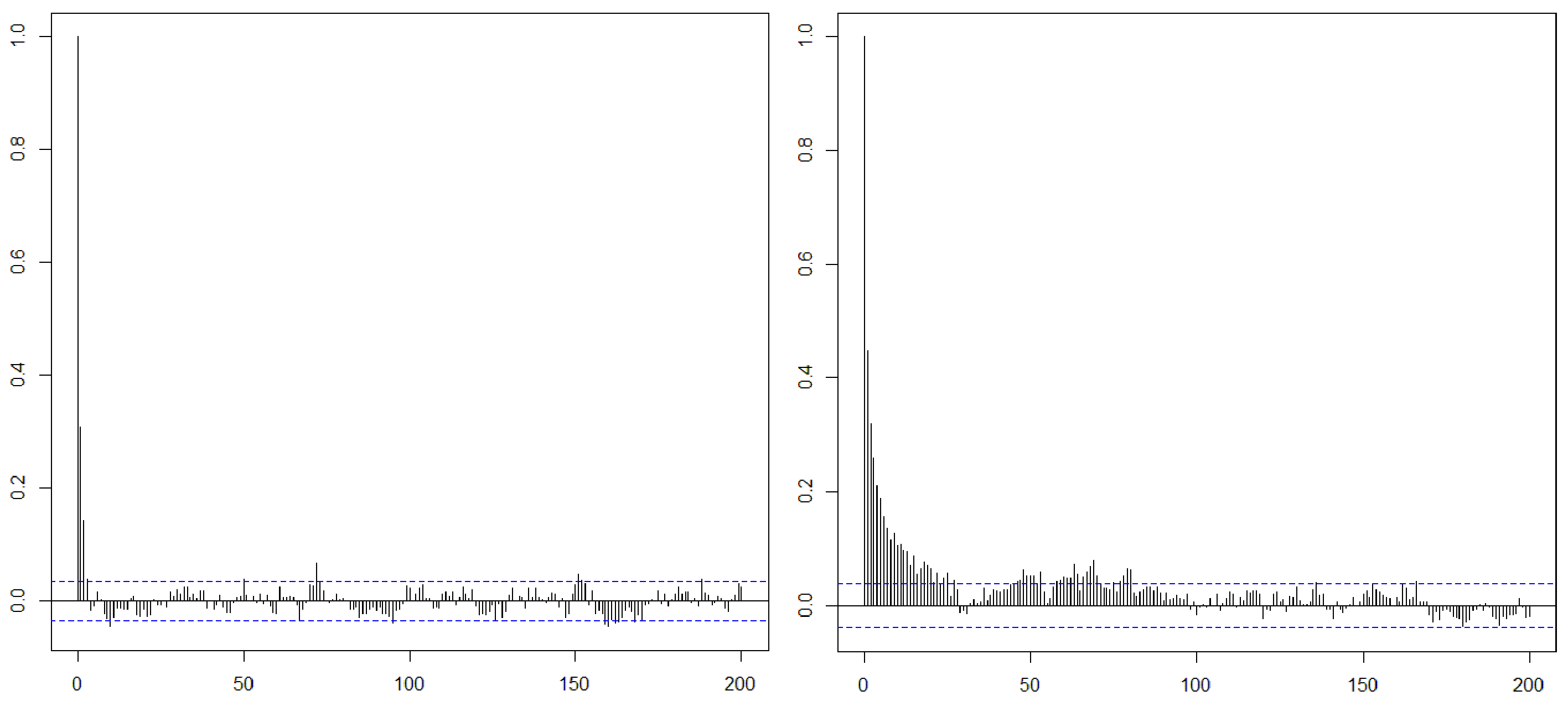}\\
\caption{ACF of $-2\log(\pi(\psi|W,S))$ for ours (left) and Adams et al.'s (right).}\label{fig4ac}
\end{figure}

Moreover, the high autocorrelation of Adams et al.'s MH subchain has great impact in the autocorrelation of the whole chain. We present the autocorrelation function of $-2\log(\pi(\psi|W,S))$ ($\pi(\psi|W,S)$ is the full posterior density) for both methodologies in the second example (see Figure \ref{fig4ac}). The inefficiency factor $1+2\sum_{l=2}^{\infty}\rho_l$ of the effective sample size was 1.38 and 12.38 for ours and Adams et al.'s models, respectively, where $\rho_l$ is the autocorrelation of order $l$ and the sum is truncated at $l=200$.

\end{document}